\documentclass{aa}

\def\het{He(2$^{3}$S)}
\def\mlr{$\dot M$}

\def\gs{g\,s$^{-1}$}
\def\rp{$R_{\rm P}$}
\newcommand{\gj}{GJ\,1214\,b}

\usepackage{subcaption}
\usepackage{graphicx}
\usepackage{txfonts}
\usepackage[colorlinks=true,citecolor=blue]{hyperref}
\begin{document}

   \title{A tentative detection of \ion{He}{I} in the atmosphere of GJ\,1214\,b}


   \author{J.~Orell-Miquel\inst{\ref{ins:iac},\ref{ins:ull}}
        \and
          F.~Murgas\inst{\ref{ins:iac},\ref{ins:ull}}
          \and
          E.~Pall\'e\inst{\ref{ins:iac},\ref{ins:ull}}
          \and
          M.~Lamp\'on\inst{\ref{ins:IAA}}
          \and
          M.~L\'opez-Puertas\inst{\ref{ins:IAA}}
          \and
          J.~Sanz-Forcada\inst{\ref{ins:jorge}}
          \and
          E.~Nagel\inst{\ref{ins:eva1},\ref{ins:eva2}}
          \and
          A.~Kaminski\inst{\ref{ins:adr1}}
          \and
          N.~Casasayas-Barris\inst{\ref{ins:nuria}}
          \and
          L.~Nortmann\inst{\ref{ins:Alem}}
          \and
          R.~Luque\inst{\ref{ins:IAA}}
          \and
          K.~Molaverdikhani\inst{\ref{ins:LMU},\ref{ins:ORIGINS},\ref{ins:adr1}} 
          \and
          E.~Sedaghati\inst{\ref{ins:IAA}}
          \and
          J.\,A.~Caballero\inst{\ref{ins:jorge}}
          \and
          P.\,J.~Amado\inst{\ref{ins:IAA}}
          \and
          G.~Bergond\inst{\ref{ins:CAHA}} 
          \and
          S.~Czesla\inst{\ref{ins:eva1},\ref{ins:eva2}} 
          \and
          A.\,P.~Hatzes\inst{\ref{ins:eva2}} 
          \and
          Th.~Henning\inst{\ref{ins:Max}} 
          \and
          S.~Khalafinejad\inst{\ref{ins:adr1},\ref{ins:Max}} 
          \and
          D.~Montes\inst{\ref{ins:UCM}} 
          \and
          G.~Morello\inst{\ref{ins:iac},\ref{ins:ull}}
          \and
          A.~Quirrenbach\inst{\ref{ins:adr1}} 
          \and
          A.~Reiners\inst{\ref{ins:Alem}}
          \and
          I.~Ribas\inst{\ref{ins:ICE},\ref{ins:IEEC}} 
          \and
          A.~S\'anchez-L\'opez\inst{\ref{ins:nuria}} 
          \and
          A.~Schweitzer\inst{\ref{ins:Hamburg}} 
          \and
          M.~Stangret\inst{\ref{ins:iac},\ref{ins:ull}} 
          \and
          F.~Yan\inst{\ref{ins:Alem}} 
          \and
          M.\,R.~Zapatero~Osorio\inst{\ref{ins:CAB2}} 
          }

   \institute{
        \label{ins:iac}Instituto de Astrof\'isica de Canarias (IAC), 38205 La Laguna, Tenerife, Spain\\
        \email{jom@iac.es}
        \and
        \label{ins:ull}Departamento de Astrof\'isica, Universidad de La Laguna (ULL), 38206 La Laguna, Tenerife, Spain
        \and
        \label{ins:IAA}Instituto de Astrof\'isica de Andaluc\'ia (IAA-CSIC), Glorieta de la Astronom\'ia s/n, 18008 Granada, Spain
        \and
        \label{ins:jorge}Centro de Astrobiolog\'{i}a (CSIC-INTA), ESAC, Camino Bajo del Castillo s/n, Villanueva de la Ca\~{n}ada, E-28692 Madrid, Spain
        \and
        \label{ins:eva1}Hamburger Sternwarte, Universit\"at Hamburg, Gojenbergsweg 112, 21029 Hamburg, Germany
        \and
        \label{ins:eva2}Th\"uringer Landessternwarte Tautenburg, Sternwarte 5, 07778 Tautenburg, Germany
        \and
        \label{ins:adr1}Landessternwarte, Zentrum f\"ur Astronomie der Universit\"at Heidelberg, K\"onigstuhl 12, 69117 Heidelberg, Germany
        \and
        \label{ins:nuria}Sterrewacht Leiden, Universiteit Leiden, Postbus 9513, 2300 RA Leiden, The Netherlands
        \and
        \label{ins:Alem}Institut f\"ur Astrophysik, Georg-August-Universit\"at, Friedrich-Hund-Platz 1, 37077 Göttingen, German
        \and
        \label{ins:LMU}Universit\"ats-Sternwarte, Ludwig-Maximilians-Universit\"at M\"unchen, Scheinerstrasse 1, D-81679 M\"unchen, Germany
        \and
        \label{ins:ORIGINS}Exzellenzcluster Origins, Boltzmannstrasse 2, 85748 Garching, Germany
        \and
        \label{ins:Max}Max-Planck-Institute f\"ur Astronomie, K\"onigstuhl 17, D-69117 Heidelberg, Germany
        \and
        \label{ins:ICE}Institut de Ci\`encies de l'Espai (ICE, CSIC), Campus UAB, Can Magrans s/n, 08193 Bellaterra, Barcelona, Spain
        \and
        \label{ins:IEEC}Institut d’Estudis Espacials de Catalunya (IEEC), 08034 Barcelona, Spain
        \and
        \label{ins:CAB2}Centro de Astrobiolog\'ia (CSIC-INTA), Carretera de Ajalvir km 4, E-28850 Torrej\'on de Ardoz, Madrid, Spain
        \and
        \label{ins:UCM}Facultad de Ciencias F\'isicas, Departamento de F\'sica de la Tierra y Astrof\'isica \& IPARCOS-UCM (Instituto de F\'isica de Part\'iculas y del Cosmos de la UCM), Universidad Complutense de Madrid, 28040 Madrid, Spain
        \and
        \label{ins:Hamburg}Hamburger Sternwarte, Gojenbergsweg 112, 21029 Hamburg, Germany.
        \and
        \label{ins:CAHA}Observatorio de Calar Alto, Sierra de los Filabres, 04550 G\'ergal, Almer\'ia, Spain
        }

   \date{Received 15 October 2021 / Accepted 12 January 2022}

 
  \abstract
  {The $\ion{He}{I}$ $\lambda$10833\,$\AA$ triplet is a powerful tool for characterising the upper atmosphere of exoplanets and tracing possible mass loss. Here, we analysed one transit of GJ\,1214\,b observed with the CARMENES high-resolution spectrograph to study its atmosphere via transmission spectroscopy around the $\ion{He}{I}$ triplet. Although previous studies using lower resolution instruments have reported non-detections of $\ion{He}{I}$ in the atmosphere of GJ\,1214\,b, we report here the first potential detection. We reconcile the conflicting results arguing that previous transit observations did not present good opportunities for the detection of $\ion{He}{I}$, due to telluric H$_2$O absorption and OH emission contamination.We simulated those earlier observations, and show evidence that the planetary signal was contaminated. From our single non-telluric-contaminated transit, we determined an excess absorption of 2.10$^{+0.45}_{-0.50}$\,\% (4.6\,$\sigma$) with a full width at half maximum (FWHM) of 1.30$^{+0.30}_{-0.25}$\,\AA. The detection of \ion{He}{I} is statistically significant at the 4.6\,$\sigma$ level, but repeatability of the detection could not be confirmed due to the availability of only one transit. By applying a hydrodynamical model and assuming an H/He composition of 98/2, we found that GJ\,1214\,b would undergo hydrodynamic escape in the photon-limited regime, losing its primary atmosphere with a mass-loss rate of (1.5--18)\,$\times$\,10$^{10}$\,g\,s$^{-1}$ and an outflow temperature in the range of 2900--4400\,K.
  Further high-resolution follow-up observations of GJ\,1214\,b are needed to confirm and fully characterise the detection of an extended atmosphere surrounding GJ\,1214\,b. If confirmed, this would be strong evidence that this planet has a primordial atmosphere accreted from the original planetary nebula. Despite previous intensive observations from space- and ground-based observatories, our $\ion{He}{I}$ excess absorption is the first tentative detection of a chemical species in the atmosphere of this benchmark sub-Neptune planet.
  } 


   \maketitle
%
\section{Introduction}

The study of atmospheric escape in exoplanet atmospheres started with the detection of excess Lyman-$\alpha$ absorption in the exosphere of HD\,209458\,b \citep{VidalMadjar2003}, demonstrating that this line is a powerful tool for probing evaporating atmospheres.
However, Ly$\alpha$ has its limitations as ($i$) it is strongly affected by interstellar extinction, limiting its application to only the closest stars due to their large relative motions, especially for K and M dwarfs that have low fluxes in the far ultraviolet (UV), and ($ii$) UV observations cannot be carried out from ground-based facilities at high spectral resolutions ($\mathcal{R}$\,$\sim$\,100\,000), relegating its study to space observations with {\em Hubble}/STIS.
At optical wavelengths, \citet{Yan_Henning_2018} proved that Balmer H$\alpha$ is also a good tracer for atmospheric evaporation. These two hydrogen lines can provide information about several mechanisms, such as photo-evaporation \citep{OwenWu2017ApJ...847...29O,JinMordasini2018ApJ...853..163J} or core-powered mass loss \citep{Ginzburg2018MNRAS.476..759G,GuptaSchlichting2020MNRAS.493..792G}, which can result in the partial or complete loss of a planet's gaseous envelope.
In particular, these processes have been suggested to explain the observed bimodal distribution of small exoplanets (e.g. \citealp{Fulton2017, Fulton_2018, VanEylen_2018}).

As an alternative to Ly$\alpha$ and H$\alpha$, \cite{Oklopcic2018} suggested observing the \ion{He}{I} triplet at 10833\,$\AA$ during a planetary transit in order to search for evidence of atmospheric escape in the infrared (IR) wavelengths. The first detections of \ion{He}{} in the atmosphere of an exoplanet occurred nearly simultaneously from space for WASP-107\,b (\citealp{Spake2018}) and from the ground for WASP-69\,b \citep{Nortmann_2018} and HAT-P-11\,b \citep{Allart_2018}. In both cases the relatively large absorption depth of the \ion{He}{} triplet was interpreted as evidence of an extended H/He-rich atmosphere escaping beyond the gravitational influence of the planet. At optical and IR wavelengths, ground-based high-resolution spectroscopy possesses great advantages over lower resolution space observations, allowing line profile determination
from which model-dependent physical parameters can be retrieved, such as the atmospheric escape rate and outflow temperature \citep{Lampon2021a}.
However, helium also presents some disadvantages: ($i$) \ion{He}{I} triplet lines are surrounded by telluric lines that can mask the planet signal, ($ii$) its ionisation wavelength cut-off ($\lambda \leq$ 504\,$\AA$) is lower than that of neutral \ion{H}{} ($\lambda \leq$ 912\,$\AA$), and ($iii$) relatively low escape rates can produce a large Ly$\alpha$ feature detectable in multiple epochs, while remaining non-detected for \ion{He}{I} (e.g. GJ\,436\,b, \citealp{Ehrenreich_2015, Lavie_2017_GJ436, Nortmann_2018, dosSantos_2019_GJ436b, Owen_2021_Lya}). Because all the line tracers have some advantages and disadvantages, observing Ly$\alpha$, H$\alpha$, and \ion{He}{I} whenever possible is important to fully understand and analyse the process of atmospheric escape from UV to IR wavelengths.

Theoretical studies of the \ion{He}{I} triplet line show that short-period planets orbiting around late-type stars (K--M) are ideal candidates to search for helium absorption (\citealp{Oklopcic2019}). However, so far, the only detections of \ion{He}{I} in a planet orbiting around an M dwarf using high-resolution transmission spectroscopy has been for the Neptune-sized planet GJ\,3470\,b \citep{Enric_2020, Ninan2020ApJ...894...97N}.
Other studies of transiting planets around M dwarfs have only placed upper limits on the presence of \ion{He}{} (e.g. GJ\,436\,b, \citealp{Nortmann_2018}; AU\,Mic\,b, \citealp{Hiranos_AUMic_He_2020}; TRAPPIST-1\,b, e, and f, \citealp{Trappist_2021}).

GJ\,1214\,b (\citealp{Charbonneau2009}) is a well-studied, benchmark, warm, Neptune-sized planet (2.7\,$\mathrm{R}_\oplus$, 8.1\,$\mathrm{M}_\oplus$), orbiting around an M4.5 dwarf with a period of 1.58\,d. Although many planets with similar properties have recently been discovered, GJ\,1214\,b was one of the first to be found with a mass and radius between those of Earth and Neptune, and it offered an exceptional opportunity to explore the composition of this type of planet with no analogues in the Solar System (in Table\,\ref{table - System Param}, we compile a comprehensive list of the system properties, either from the literature or derived in this study). Thus, this planet has been the subject of intense observing campaigns from space- and ground-based observatories, revealing a cloudy atmosphere with a featureless transmission spectrum (\citealp{Bean2010,Berta_2012_Flat_spectrum, Kreidberg_2014}).
In particular, there have been several attempts to detect or place upper limits on the presence of \ion{He}{I} in the atmosphere of this planet (e.g. \citealp{Crossfield2019, delaRoche2020, Kasper_2020}), although so far a detection has not been achieved.
Here, we report a tentative detection of \ion{He}{I} in the upper atmosphere of GJ\,1214\,b using high-resolution transmission spectra taken with the CARMENES spectrograph.

\section{Observations and data analysis}

\begin{table}
\caption{\label{table - System Param} Stellar and planet parameters of the GJ\,1214 system.
}
\centering
\resizebox{\columnwidth}{!}{%
\begin{tabular}{lcl}
\hline
\hline
\noalign{\smallskip}
Parameter & Value & Reference \\
\noalign{\smallskip}
\hline
\noalign{\smallskip}

\multicolumn{3}{c}{\textit{Stellar parameters}} \\
$M_{\star}$ [$M_{\odot}$] & 0.150 $\pm$ 0.011 & H13 \\
$R_{\star}$ [$R_{\odot}$] & 0.216 $\pm$ 0.012 & H13 \\
$T_{\mathrm{eff}}$ [K] & 3026 $\pm$ 150 & H13 \\
$L_{\rm X}$ [erg s$^{-1}$] & $7.7 \times 10^{25}$ & Sect.\,\ref{Sec:XMM} \\
$d$ [pc] & 14.65 $\pm$ 0.03 & \textit{Gaia}\,DR2$^{a}$ \\

\noalign{\smallskip}
\multicolumn{3}{c}{\textit{Planet parameters}}\\
\noalign{\smallskip}

$M_{p}$ [$M_{\oplus}$] & 8.14 $\pm$ 0.43 & C21 \\
$R_{p}$ [$R_{\oplus}$] & 2.74 $\pm$ 0.05 & C21 \\
$\rho_{p}$ [g\,cm$^{-3}$]  & 2.20 $\pm$ 0.17 & C21 \\
$K_p$ [km\,s$^{-1}$] & 97.1 $\pm$ 2.2 & This work$^{b}$ \\
$T_{\mathrm{eq}}$ [K] & 596 $\pm$ 19 & C21 \\

\noalign{\smallskip}
\multicolumn{3}{c}{\textit{Transit and system parameters}}\\
\noalign{\smallskip}

$P$ [d] & 1.58040418\,(24) & C11 \\
$T_0$ [BJD] & 2454980.748996\,(45) & C11 \\
$T_{14}$ [min] & 52.73$^{+0.49}_{-0.35}$ & C11$^{c}$ \\
$T_{12}$ [min] & 5.80$^{+0.41}_{-0.45}$ & C11$^{c}$ \\
$a$ [au] & 0.01411\,(32) & H13 \\
$i_p$ [deg] & 88.17 $\pm$ 0.54 & H13 \\
$\gamma$ [km\,s$^{-1}$] & +21.1 $\pm$ 1.0 & C09$^{d}$ \\
$K_{\star}$ [m\,s$^{-1}$] & 14.36 $\pm$ 0.53 & C21 \\

\noalign{\smallskip}
\hline
\end{tabular}
}

\tablebib{H13: \cite{SysPara_Harpsoe}, \textit{Gaia}\,DR2: \cite{GAIA_DR2}, L14: \citet{lal14}, C21: \citet{Cloutier_2021}, C09: \cite{Charbonneau2009}, C11: \cite{SysPara_Carter2011}.}

\tablefoot{
$^{(a)}$ There is a more recent and precise parallactic distance value from {\em Gaia} EDR3 at $d = 14.642 \pm 0.014$\,pc, but we kept the DR2 value for a better comparison with previous works.
$^{(b)}$ Calculated from $K_p = 2 \pi a P^{-1} \sin{i_p}$ using the parameters in this table.
$^{(c)}$ $T_{14}$ is the total transit duration between the fist ($T_1$) and fourth ($T_4$) contacts, and $T_{12}$ is the duration of the ingress between the first and second contacts.
$^{(d)}$ We corrected the system velocity ($\gamma$) sign from \citet{Charbonneau2009} as \citet{Kasper_2020} remarked.
}

\end{table}

\subsection{CARMENES observations}
\label{Sec:Observations}

A single transit of GJ\,1214\,b was observed with the Calar Alto high-Resolution search for M dwarfs with Exoearths with Near-infrared and optical \'Echelle Spectrographs (CARMENES; \citealp{Quirrenbach_2014, Quirrenbach_2020}), located at the Calar Alto Observatory, Almer\'ia, Spain, on the night of 4 May 2021. CARMENES has two spectral channels: the optical channel (VIS), which covers the wavelength range from 0.52 to 0.96\,$\mu$m, and the near-IR channel (NIR), which goes from 0.96 to 1.71\,$\mu$m. Although we observed the target with both channels simultaneously, here we report results from the analysis of spectra with the NIR one, where the \ion{He}{I} triplet is located.

We collected a total of 27 high-resolution NIR spectra ($\mathcal{R}$\,=\,80\,400), seven of them between the first ($T_1$) and fourth ($T_4$) contacts.
The spectra were taken in good weather conditions with an exposure time of 358\,s, ensuring that 
any planetary absorption line was not spread over more than $\sim$\,2 pixels during any given exposure, and with a signal-to-noise ratio (S/N) that varied from 18 to 35 (median value of 28) around 10\,830\,$\AA$.
Fibre A was used to observe GJ\,1214, while fibre B was placed on sky in order to monitor the sky emission lines (fibres A and B are permanently separated by 88\,arcsec in the east-west direction).
The observations were reduced using the CARMENES pipeline \texttt{caracal} \citep{Caballero2016}, and both fibres were extracted with the flat-optimised extraction algorithm (\citealp{FOX_extraction}).

\subsection{{\em XMM-Newton} \emph{and} {\em Hubble} \emph{observations}}
\label{Sec:XMM}

X-ray and UV spectra\footnote{The spectrum is publicly available in the CARMENES data archive (\url{http://carmenes.cab.inta-csic.es/}) and will also be published in X-Exoplanets (\url{http://sdc.cab.inta-csic.es/xexoplanets/jsp/homepage.jsp}).} were used to provide the planet atmospheric model with the spectral energy distribution of the star in high energies (see Sect.\,\ref{Sec:model}). {\em XMM-Newton} observations (P.I. Lalitha) of GJ\,1214 were taken on 27 September 2013. An X-ray luminosity of $L_{\rm X} \simeq$ 7.4$\times$10$^{25}$\,erg\,s$^{-1}$ was measured in the EPIC spectra \citep{lal14}. We re-analysed the same data in order to model the X-ray (5--100\,$\AA$) and extreme UV (EUV, 100--920\,$\AA;$ hereafter XUV 5--920\,$\AA$), and spectral emission of the star. The target is detected with S/N\,=\,3.7 considering the three EPIC cameras (pn, MOS1, MOS2). A global-fit model with a single temperature was used to jointly fit the three EPIC spectra. The spectra were fit using the \texttt{ISIS} \citep{isis} software and the atomic data from the Astrophysical Plasma Emission Database \citep[APED,][]{aped}. Given the poor statistics, we needed to fix the values of coronal abundance, and we assumed an interstellar medium absorption of $N_{\rm H}$\,=\,3$\times$10$^{18}$\,cm$^{-3}$, which is appropriate for its distance \citep[$d$\,=\,14.65\,pc,][]{2003ApJS..145..147S}. The resulting model has a temperature of $\log{T\,[{\rm K}]}\,=\,6.56^{+1.3}_{-0.16}$ and an emission measure ($EM$) of $\log{EM\,[{\rm cm}^{-3}}]\,=\,48.43^{+0.16}_{-0.25}$. Our analysis of the EPIC spectra indicates a flux of $L_{\rm X} \simeq$ 7.7$\times$10$^{25}$\,erg\,s$^{-1}$, which is consistent with the value previously reported.

Ultraviolet COS and STIS observations of GJ\,1214 were collected from the {\em Hubble Space Telescope} Legacy archive. {\em Hubble}/COS observations were performed with the G130M and G160M grating, while {\em Hubble}/STIS spectra were collected using the G230L grating. All {\em Hubble} observations were collected on 24 August 2012 and from the 19--21 August 2015, and they were made available through the MUSCLES programme \citep{you16}. The spectra have poor statistics, with the error level being higher than the observed continuum in most of the spectral range. This is expected since most K and M dwarfs have low continuum fluxes in the UV range. Few lines were measured with S/N\,$>$\,3 and were used to calculate the emission measure distribution in the low temperature range following the procedures described in \citet{san11}.

The X-ray $EM$ and $T$ values were used to complement a coronal model
at high temperatures (Sanz-Forcada, priv comm.). The modelled emission indicates an XUV luminosity of 3.6$\times$10$^{26}$\,erg\,s$^{-1}$ in the 5--920\,$\AA$ spectral range, and 1.7$\times$10$^{26}$\,erg\,s$^{-1}$ in the 5--504\,$\AA$ region. 
These values are lower than those reported by \citet{lal14}, who scaled the EUV luminosity from the X-ray value.


\subsection{Telluric contamination removal}
\label{Sec:Telluric}

\begin{figure}
    \centering
    \includegraphics[width=\hsize]{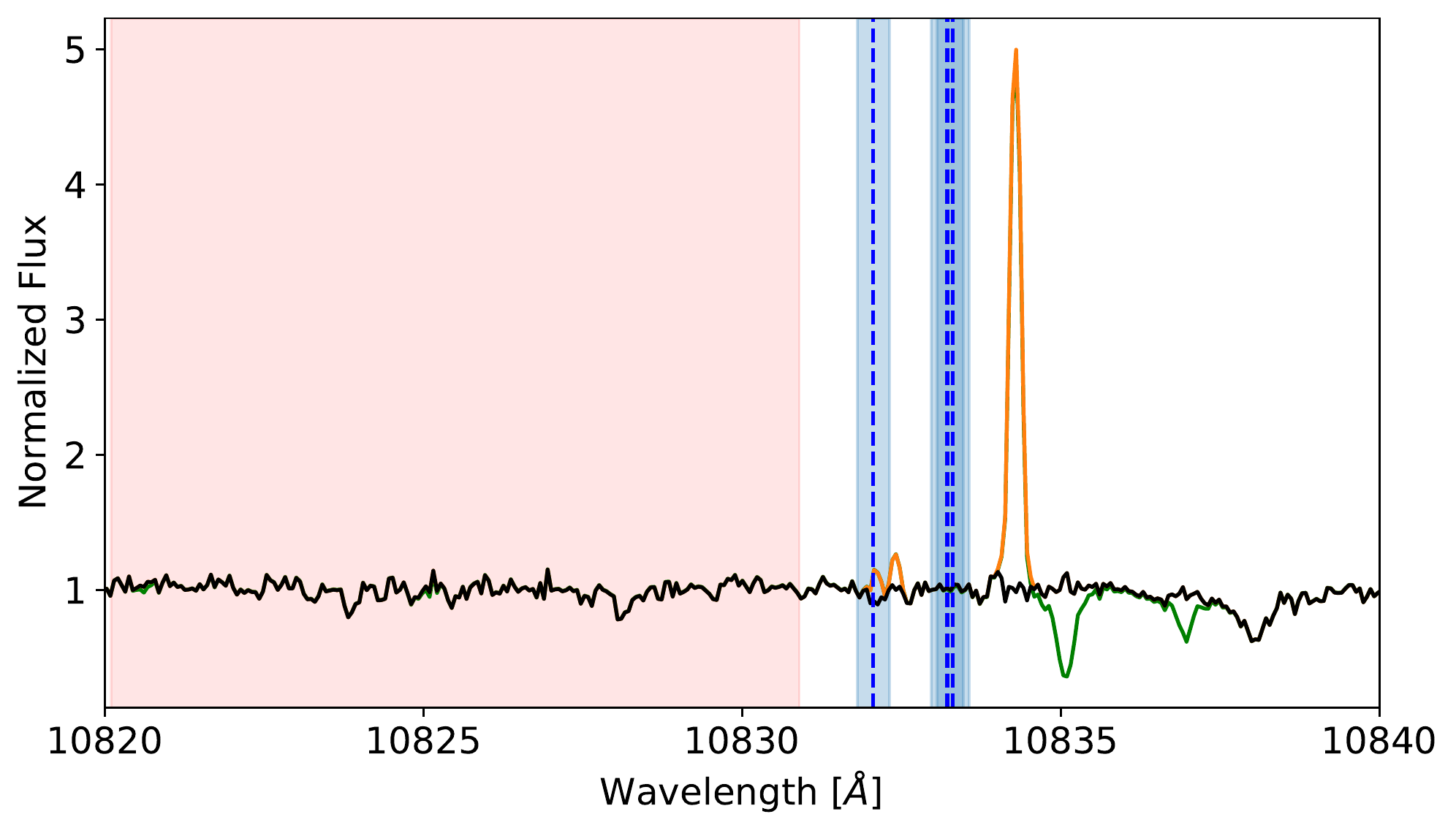}
    \caption{Details of one CARMENES spectrum of GJ\,1214 around the \ion{He}{I} triplet lines. A raw spectrum after standard data reduction is plotted in green. The same spectrum after the telluric correction with \texttt{molecfit} is over-plotted in orange. The same spectrum after removing the OH emission features is over-plotted in black.
    The red shaded region is the spectral range used to normalise the continuum of all spectra.
    The vertical blue dashed lines indicate the position at mid-transit of \ion{He}{I} triplet lines and the blue shaded regions represent the planet traces during the transit.
    All the spectra are shifted to the stellar rest frame at vacuum wavelength.
    }
    \label{Fig: Tell Corr}
\end{figure}

Figure\,\ref{Fig: Tell Corr} shows that surrounding the \ion{He}{I} triplet lines, where the planetary absorption is expected, there are H$_2$O absorption and OH emission lines from the Earth's atmosphere.
Because GJ\,1214's system velocity is +21.1\,km\,s$^{-1}$ and the barycentric Earth radial velocity (BERV) oscillates approximately between $+$27\,km\,s$^{-1}$ and $-$27\,km\,s$^{-1}$, the bluest and reddest telluric shifts are not enough to not affect the regions of interest (Fig.\,\ref{Fig: MIN_MAX_OH_position}). In between these two opposite configurations, one always finds some degree of telluric contamination, and their effects need to be accounted for. Our observations were planned to avoid a complete overlap or superposition of the telluric lines and the \ion{He}{I} planetary trace. We mitigated the contamination by observing at an epoch when the faintest lines of the OH emission overlapped only the faintest line of the \ion{He}{I} triplet.

We dealt with this telluric contamination following two different methods.
The first strategy was to simply mask the regions of the spectra affected by telluric emission and absorptions.
In that case, when we shifted the spectra to the planet rest frame, some particular wavelengths were represented by fewer pixels, decreasing the S/N of the final transmission spectrum \citep{Enric_2020, Nuria_WASP76}.

In the second strategy, we corrected the telluric H$_2$O absorption and OH emission lines using synthetic-line models.
The spectra obtained were firstly corrected for telluric absorption molecules that are present in the CARMENES NIR spectrum, viz. H$_2$O, O$_2$, CH$_4$, and CO$_2$, using the \texttt{molecfit} package in version 1.5.9 (\citealp{molecfit_1,molecfit_2}). The initial \texttt{molecfit} fit parameters, as well as the fitted wavelength ranges, are summarised in Tables\,\ref{table - molecfit parameters} and~\ref{table - molecfit ranges}. Then, we corrected for OH telluric emission following the methodology applied and described in previous \ion{He}{I} studies (\citealp{Nortmann_2018, Salz_2018, AlonsoFloriano_2019, Enric_2020, Nuria_WASP76, Czesla_2021}), where CARMENES fibre B, which is pointed at the sky during observations, is used to generate an OH emission model to correct it in the science spectra.
In each sky spectrum, we simultaneously fitted the three main OH emission features near 10833\,\AA\ with three independent Gaussian profiles.
We set the amplitude, central position and standard deviation free, but with an initial guess for the central positions (10832.1\,\AA, 10832.4\,\AA,\ and 10834.25\,\AA).
Although the spectra from both fibres were extracted with the same procedure, they have different amplitudes due to a difference in efficiencies of the two injection fibres.
To take care of this, we computed a scaling factor between the two fibres from a high S/N spectrum of each fibre.
For each pair of sky and target spectra, we subtracted the particular sky emission model scaled by the efficiency factor (0.906; assumed constant during the night) from fibre A.
Using this strategy, some residuals may remain, contaminating the final transmission spectra.
The effect of telluric absorption and emission removals is illustrated in Fig.\,\ref{Fig: Tell Corr}. For a more detailed explanation of the telluric emission contamination removal, we invite the reader to consult Sect.\,B in \citet{Czesla_2021}.

The next step after the telluric absorption and emission removal was the normalisation of all spectra. We considered the mean value of the spectral region between 10820.1\,\AA\ and 10830.9\,\AA, which is almost unaffected by telluric absorption and near the lines of interest (see Fig.\,\ref{Fig: Tell Corr}). Finally, we computed the transmission spectrum around the \ion{He}{I} triplet lines, at 10832.06\,$\AA$, 10833.22\,$\AA$, and 10833.31\,$\AA$ (vacuum wavelength), following the well-established procedure presented in \citet{Wyttenbach_2015} and \citet{Nuria_2017}.
To compute the transmission spectra, we assumed the system and stellar parameters given in Table\,\ref{table - System Param} and combined the seven residual in-transit spectra after shifting them to the planetary rest frame.

\section{Results and discussion}
\label{Sec:Results}

\begin{figure*}
   \centering
   \includegraphics[width=\textwidth]{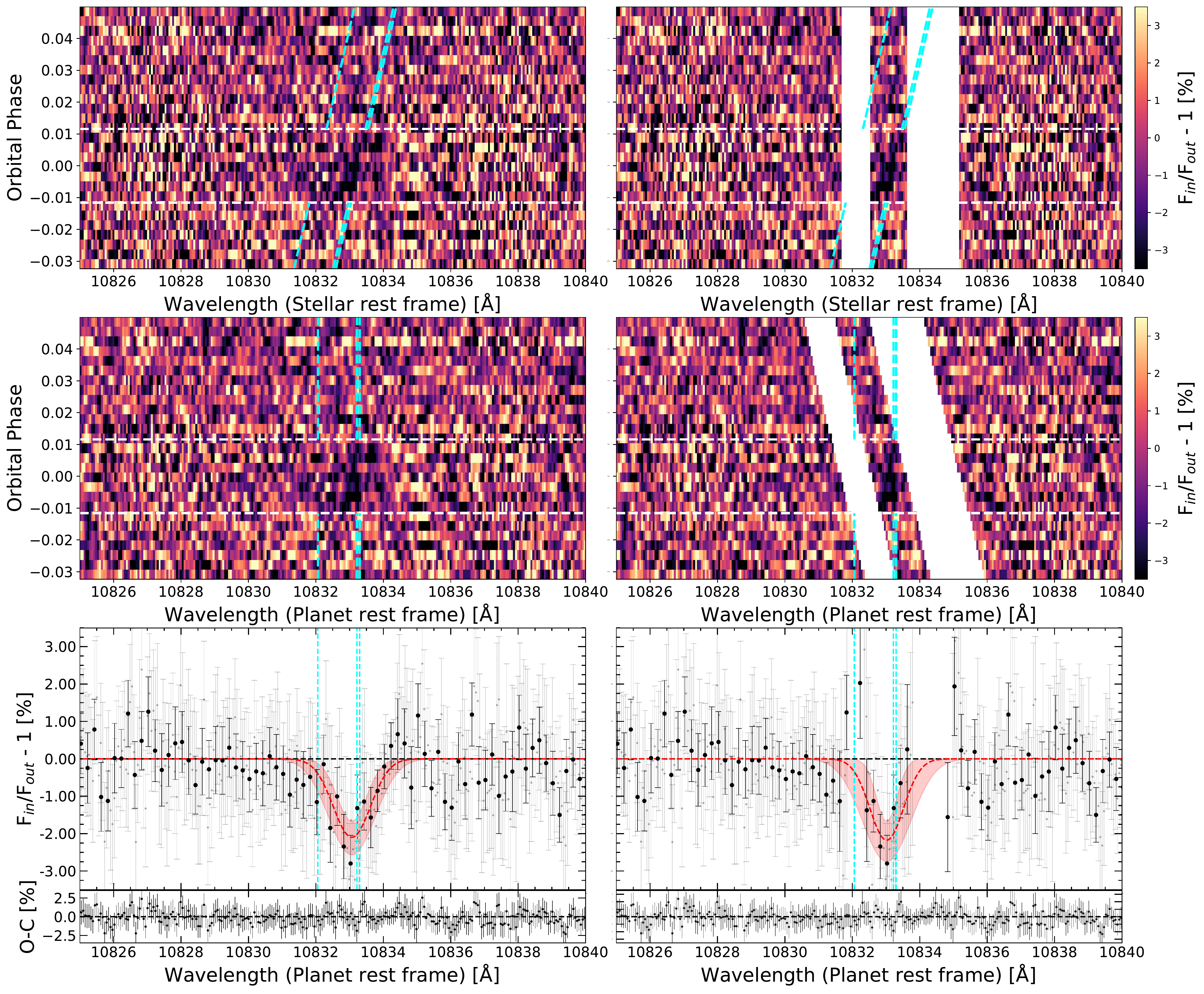}
   \caption{
        Residual maps and transmission spectra around the \ion{He}{I} triplet lines for the corrected (\textit{left column}) and masked (\textit{right column}) procedures.
        \textit{Top panels}: Residual maps in the stellar rest frame.
        \textit{Middle panels}: Residual maps in the planet rest frame.
        Planet orbital phase is shown in the vertical axis, wavelength is in the horizontal axis, and relative absorption is colour-coded.
        Dashed white horizontal lines indicate the first (T1) and fourth (T4) contacts.
        Cyan lines show the theoretical trace of the planetary signals.
        \textit{Bottom panels}: Transmission spectra obtained around an \ion{He}{I} triplet, combining all the spectra between T1 and T4. We show the original data in light grey and the data binned by 0.2\,\AA\ in black. Cyan vertical lines indicate the \ion{He}{I} triplet lines' positions. Best Gaussian fit model is shown in red along with its $1\sigma$ uncertainties (shaded red region). Residuals after subtracting the best fit model are shown below. All the wavelengths in this figure are referenced in the vacuum.
   }
    \label{Fig:MAPS_TS}
\end{figure*}

\subsection{\ion{He}{i} transmission spectrum}

\begin{figure*}
   \centering
   \begin{subfigure}{0.49\textwidth}
         \centering
         \includegraphics[width=\textwidth]{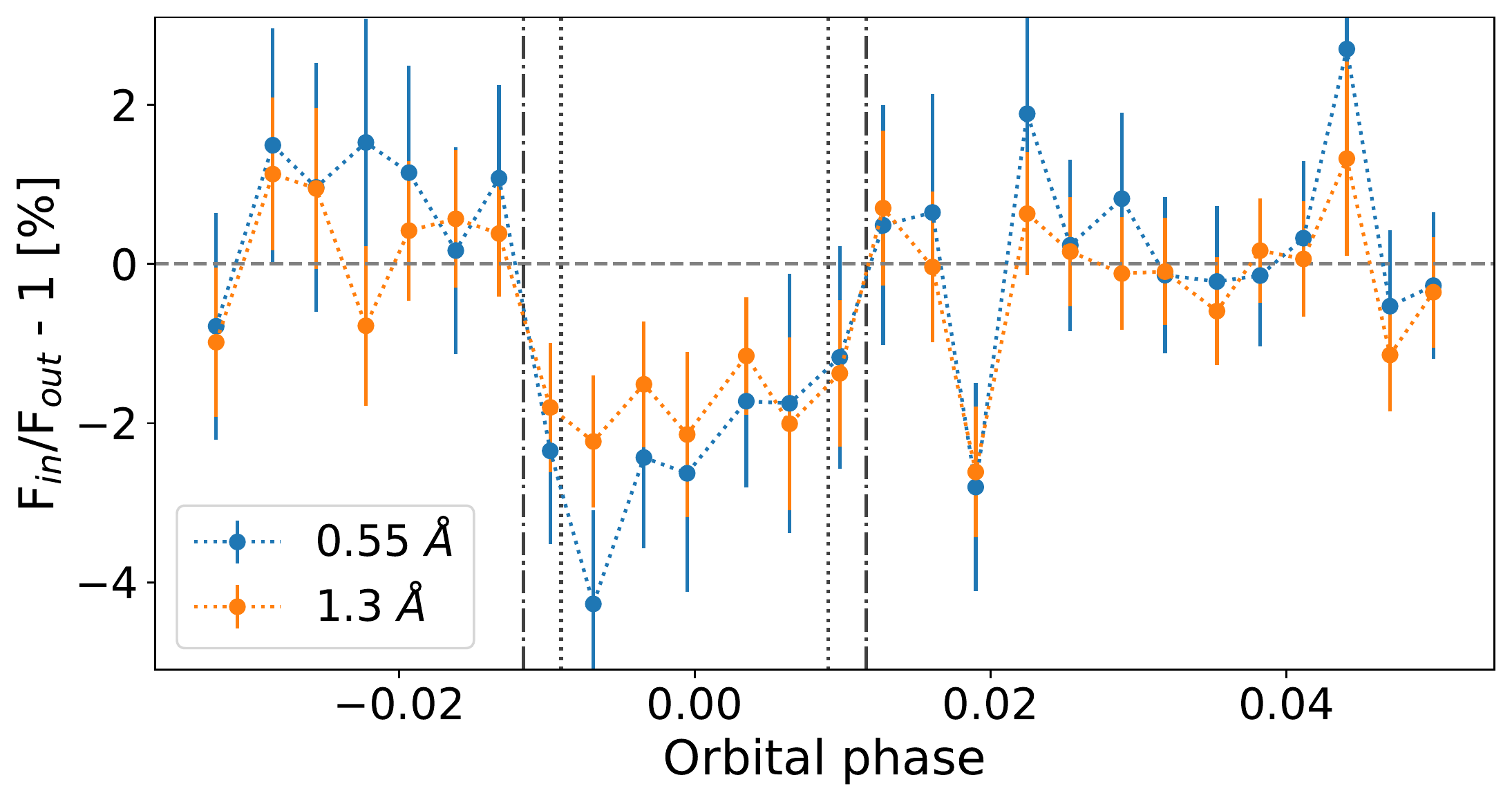}
     \end{subfigure}
     \hfill
     \begin{subfigure}{0.49\textwidth}
         \centering
         \includegraphics[width=\textwidth]{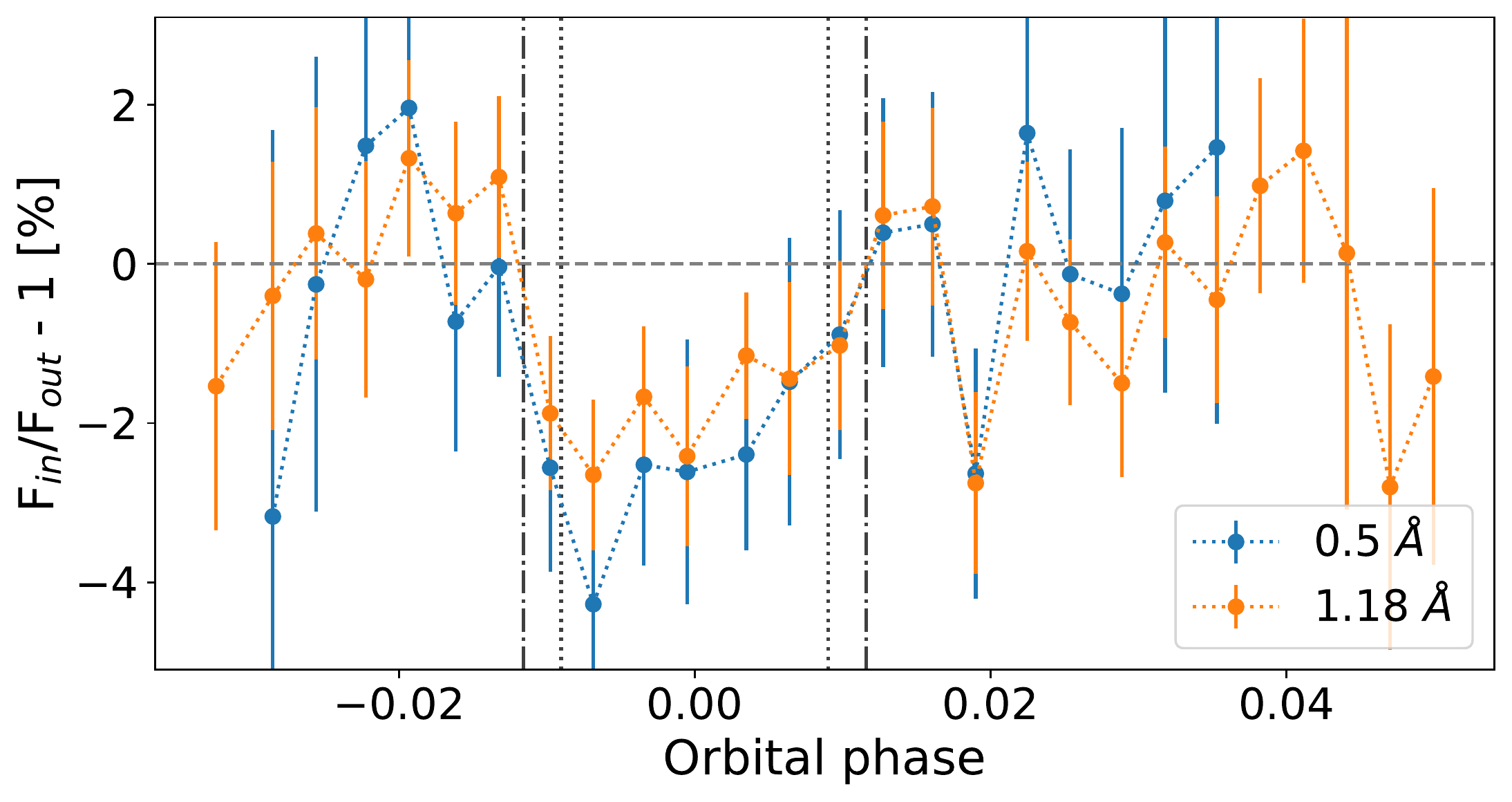}
     \end{subfigure}
   \caption{ Transit light curves of the \ion{He}{I} triplet absorption for the corrected (\textit{left}) and masked (\textit{right}) procedures. The light curves have been constructed integrating the counts around $\lambda_0$ using $\sigma$ (blue) and FWHM (orange) wavelength band passes from Table\,\ref{table - Results}. The vertical lines represent the different contacts during the transit.}
    \label{Fig: TLC}
\end{figure*}

\begin{table}
\caption{\label{table - Results} Best-fit values of the Gaussian parameters for the corrected and masked transmission spectra$^a$. }
\centering
\begin{tabular}{lcc}
\hline
\hline
\noalign{\smallskip}
Parameter & Corrected & Masked \\
\noalign{\smallskip}
\hline
\noalign{\smallskip}
Absorption [\%] & $-2.10^{+0.45}_{-0.50}$ & $-2.2^{+0.5}_{-0.6}$ \\
    \noalign{\smallskip}
$\lambda_0$ [\AA] & 10833.07$^{+0.14}_{-0.16}$ & 10833.02$^{+0.16}_{-0.25}$ \\
    \noalign{\smallskip}
$\sigma$ [\AA] &  0.55$^{+0.13}_{-0.11}$ & 0.50$^{+0.22}_{-0.20}$ \\
    \noalign{\smallskip}
FWHM [\AA] & 1.30$^{+0.30}_{-0.25}$ & 1.18$^{+0.50}_{-0.45}$ \\
    \noalign{\smallskip}
$\Delta v$ [km\,s$^{-1}$] & $-$4\,$\pm$4 & $-$5$^{+4}_{-7}$ \\
\noalign{\smallskip}
\hline
\end{tabular}
\tablefoot{
$^{(a)}$ Velocity shift $\Delta v$ is relative to the \ion{He}{I} $\lambda$10833.22\,$\AA$ line.
}
\end{table}

Figure\,\ref{Fig:MAPS_TS} shows the residual maps and the transmission spectra\footnote{The corrected transmission spectrum is publicly available in the CARMENES data archive (\url{http://carmenes.cab.inta-csic.es/}).} following the two methodologies applied to deal with the telluric contamination.
Both transmission spectra show similar excess absorption features around the two strongest lines of the \ion{He}{I} triplet.
To measure this absorption, we fitted the visible planetary signal with a Gaussian profile, sampling from the parameter posterior distributions using the MultiNest algorithm (\citealp{MultiNest}) via its {\tt python} implementation \texttt{PyMultinest} \citep{PyMultiNest}.
We used uniform priors for the excess absorption between $-10$\,\% and 0\,\% and the central position ($\lambda_0$) from 10831\,\AA\ to 10835\,\AA.
Table\,\ref{table - Priors} summarises the priors used.

Table\,\ref{table - Results} shows the results from the nested sampling fit and Figure\,\ref{Fig: Corner plot} presents the corner plots of the posterior distributions.
Although the values from the masked transmission spectrum have systematically larger errors due to their lower S/N, the results from both approaches are consistent within their $1\sigma$ uncertainties.
From the corrected transmission spectrum, we derived an absorption of $-$2.10$^{+0.45}_{-0.50}$\,$\%$ (a 4.6\,$\sigma$ level detection), a full width at half maximum (FWHM) of 1.30$^{+0.30}_{-0.25}$\,\AA,\ and a net shift of $-$4\,$\pm$4\,km\,s$^{-1}$.
The masked transmission spectrum shows an absorption of $-2.2^{+0.5}_{-0.6}$\,$\%$ (a 4.4\,$\sigma$ level detection), a FWHM of 1.18$^{+0.50}_{-0.45}$\,\AA,\ and a net shift of $-$5$^{+4}_{-7}$\,km\,s$^{-1}$.

\begin{table}
\caption{\label{table - TLC} Absorption depths retrieved for the transit light curves$^a$.}
\centering
\begin{tabular}{lccc}
\hline
\hline
\noalign{\smallskip}
Procedure & BP$_\sigma$ & BP$_{\rm FWHM}$ & TS \\
\noalign{\smallskip}
\hline
\noalign{\smallskip}
Corrected & $-2.3$\,$\pm$\,0.5 & $-1.75$\,$\pm$\,0.35 & $-2.10^{+0.45}_{-0.50}$ \\
    \noalign{\smallskip}
Masked & $-2.40$\,$\pm$\,0.55 & $-1.7$\,$\pm$\,0.4 & $-2.2^{+0.5}_{-0.6}$ \\
\noalign{\smallskip}
\hline
\end{tabular}
\tablefoot{
$^{(a)}$ Comparison of absorption depths, in percentages, retrieved for the transit light curves from Fig.\,\ref{Fig: TLC} between $T_1$ and $T_4$ and the absorption from the transmission spectra (TS). For each of the telluric correction procedures, we computed the averaged absorption over two band-passes. BP$_\sigma$ and BP$_{\rm FWHM}$ are equal to the fitted $\sigma$ and FWHM widths from Table\,\ref{table - Results}, respectively.
}
\end{table}

The transit light curve (TLC) of individual lines were useful to explore if the absorption features reported in the transmission spectra have a duration compatible with the planetary transit.
Thus, we computed the TLC for the \ion{He}{i} triplet line following the methodology explained in detail by \citet{Nortmann_2018} and \cite{Nuria_2017}. We integrated the counts in the planet rest frame for two band-passes (BP$_\sigma$ is equal to the fitted $\sigma$ width and BP$_{\rm FWHM}$ is equal to the FWHM width) centred at $\lambda_0$ for each procedure.

Figure\,\ref{Fig: TLC} shows the TLC for each methodology applied to deal with the telluric contamination and for the two different band-passes and Table\,\ref{table - TLC} compares the TLC retrieved depths along the fitted absorption from the transmission spectra.
The error bars take into account the individual scatter of each spectrum and the number of points integrated and, as in the transmission spectra, the TLC from the masked data shows the larger errors.
The four TLC exhibit a flux decrease stable and coincident with the expected first and fourth transit contact times.
The retrieved depths between first and fourth contacts, both considering the two band passes, were 1$\sigma$ consistent with that retrieved from the transmission spectrum analyses.

The TLCs did not present extra absorption extending before the first contact and further than the fourth contact.
We calculated the expected signal during the pre- and post-transit using the model presented in Section\,\ref{Sec:model}.
The computed excess absorptions were just at the noise level of the transmission spectrum, so it is consistent with no significant detection out of transit.
Further observations will allow us to increase the S/N and refine the spectro-photometric results.

\begin{figure}
    \centering
    \includegraphics[width=\hsize]{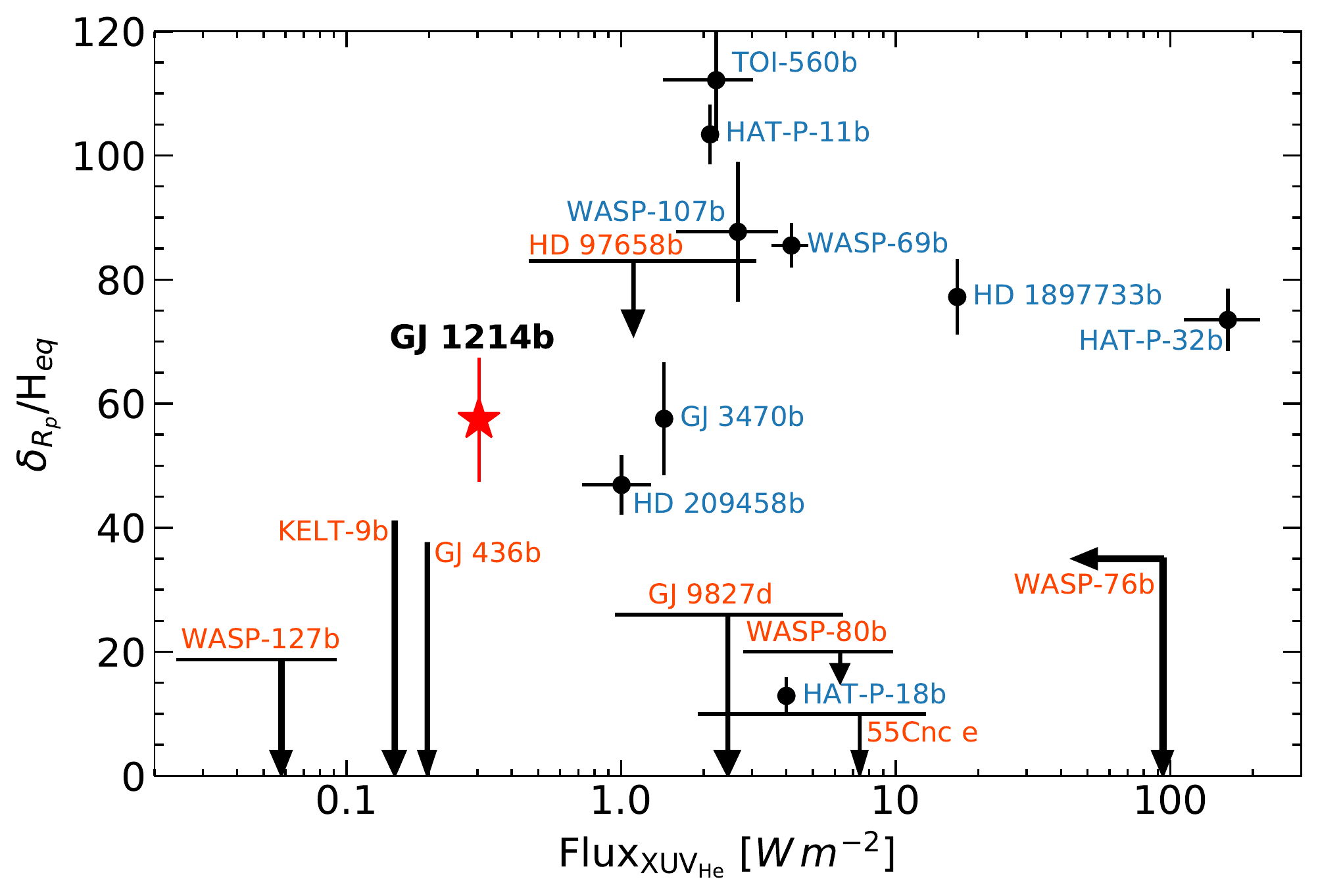}
    \caption{ Transmission signals for the planets with \ion{He}{I} detections (black dots with error bars labelled in blue) and some planets with upper limits (black arrows labelled in orange), as a function of the stellar F$_{\mathrm{XUV}_{\mathrm{He}}}$ ($\lambda$\,=\,5\,--\,504\,\AA) at the planet distance. TOI-560\,b, WASP-80\,b, and 55\,Cnc\,e flux values were computed up to $\lambda$\,$\sim$\,912\,$\AA$.
    We show the equivalent height of the \ion{He}{I} atmosphere, $\delta_{\mathrm{R_p}}$, normalised by the atmospheric scale height, $H_{\mathrm{eq}}$, of the respective planet. The red star is the results presented here on GJ\,1214\,b.
    Data on other planets are from \citet{Nortmann_2018, AlonsoFloriano_2019, Enric_2020, Kasper_2020, dosSantos_2020, Paragas_2021_HAT-P-18b, Nuria_WASP76, Zhang_2021_TOI-560b, Zhang_2021_55Cnce, Fossati_2021, Czesla_2021}. }
    \label{Fig: Planets}
\end{figure}

We computed the equivalent height of the \ion{He}{I} atmosphere ($\delta_{\mathrm{R_p}}$) normalised by its scale height ($H_{\mathrm{eq}}$; assuming $\mu$\,=\,2.3) obtaining $\delta_{\mathrm{R_p}}$/$H_{\mathrm{eq}}$\,=\,57$\pm$10.
This value is close to the one derived for GJ\,3470\,b, for which \citet{Enric_2020} also detected an \ion{He}{I} excess absorption of 1.5\,$\pm$\,0.3\,$\%$.
Figure\,\ref{Fig: Planets} puts the absorption from this work into context with regard to other \ion{He}{I} detections and some upper limits.

\subsection{Comparison with previous studies}

\begin{figure*}
   \centering
   \includegraphics[width=\textwidth]{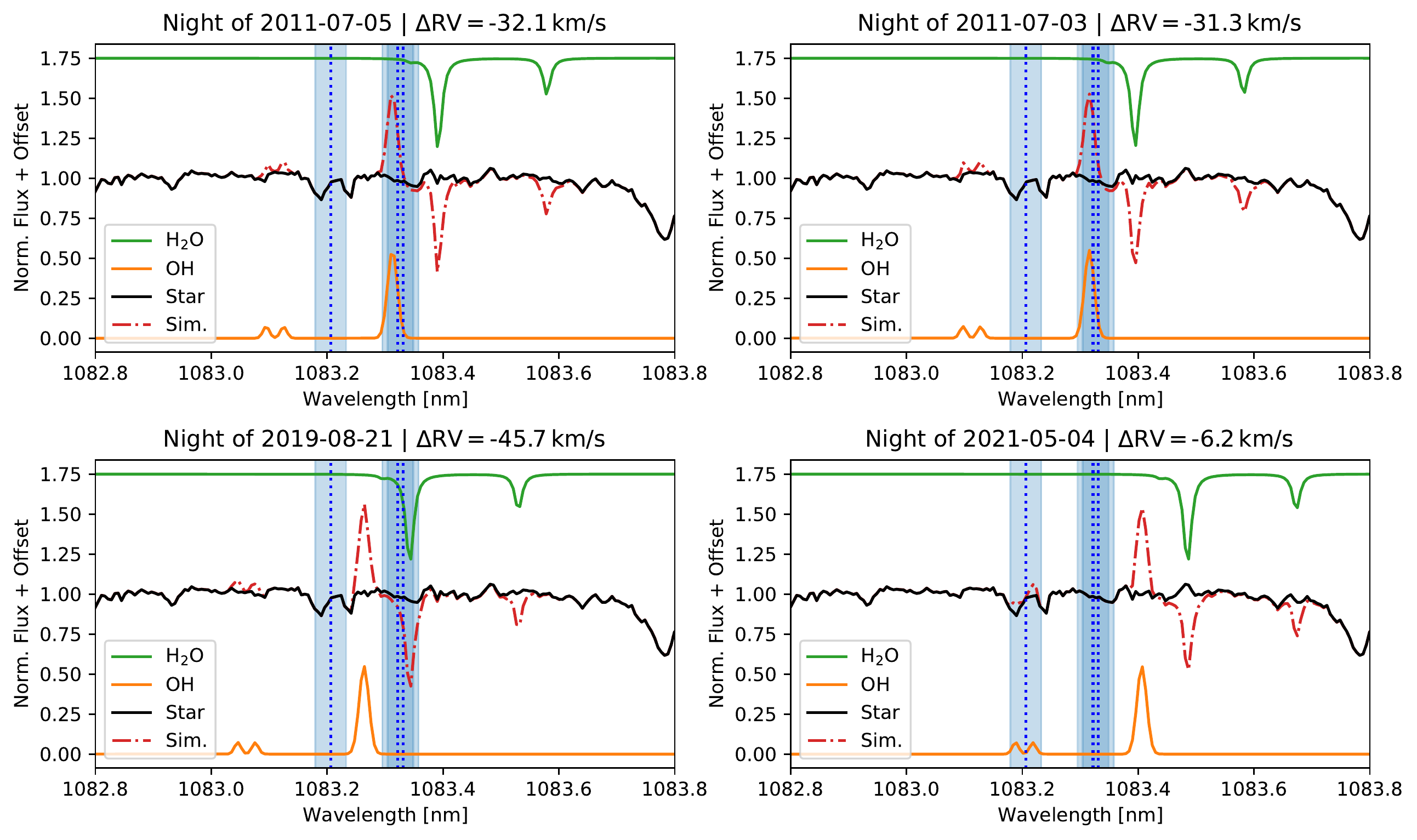}
   \caption{Simulation of the contamination of the spectrum of GJ\,1214 by H$_2$O absorption and OH emission during the nights corresponding to the observations by \citet{delaRoche2020} (2011 July 5, 2011 July 3), \citet{Kasper_2020} (2019 August 21), and this work (2021 May 4). Observing epoch and relative radial velocity between the Earth and GJ\,1214 ($\Delta$RV) are displayed above each panel. The green curve is a synthetic model of H$_2$O absorption. The orange curve is a synthetic model of OH emission. The black curve is the average of normalised out-of-transit GJ\,1214 spectra. The dashed red line is the combination of synthetic telluric models and the spectrum of GJ\,1214. The vertical blue dotted lines indicate the position of the \ion{He}{I} triplet lines and the blue shaded region represents the planet trace in the stellar rest frame at vacuum wavelength.
   We note that \ion{He}{I} lines and H$_2$O telluric absorption positions differ between this figure and Fig.\,2 from \citet{Kasper_2020} due to an error in their plot (D.\,Kasper, priv. comm.).
   }
    \label{Fig: Telluric}
\end{figure*}

GJ\,1214\,b has already been studied by other research groups through spectroscopic observations, although at significantly lower spectral resolution than the data analysed in this work.
\citet{Crossfield2019} observed GJ\,1214\,b using IRTF/SpeX in SXD mode with a spectral resolution of $\mathcal{R}$\,$\simeq$\,500.
From their non-detection of \ion{He}{I}, they estimated an excess depth of 3.8\,$\pm$\,4.3\,\% for high-resolution spectrographs such as CARMENES. This upper limit is consistent with the results presented here.

However, our detected absorption is difficult to reconcile with the refined upper limits reported in \citet{Kasper_2020} using Keck/NIRSPEC with a resolving power of $\mathcal{R}$\,$\simeq$\,25\,000 and \citet{delaRoche2020} using archival data from VLT/X-Shooter with a spectral resolution of $\mathcal{R}$\,$\simeq$\,2\,600.
Figure\,5 from \citet{Kasper_2020} does not show excess absorption features in the GJ\,1214\,b transmission spectrum above the noise level ($\sim$0.25\,\%), and \citet{delaRoche2020} reported a non-detection excess depth of 0.38\,$\pm$\,0.44\,\%.
Both results set upper limits to the absorption that are considerably lower than our detection and deviate by more than $3\sigma$ from our excess absorption ($\sim$2.1\,\%).
The main difference with these observations lies in the instrumentation used and the epoch when were carried out.
Our observations were performed with a fibre-fed, high-resolution spectrograph, in contrast to the non-detection analyses that made use of slit-fed spectrographs with significantly lower resolving power. However, more important perhaps is the difference in telluric contamination effects between the different epochs.

The wavelength region around the \ion{He}{i} triplet is affected by nearby telluric emission and absorption lines that can mask the faint planetary signal.
The relative position between these lines and those from the exoplanet atmosphere depends on the relative radial velocity between the 
BERV and stellar systemic radial velocity, which we denote here as $\Delta$RV.
This relative position varies during the year and should be taken into account when \ion{He}{I} observations are planned to avoid the overlap of tellurics and the planetary trace. The GJ\,1214\,b transit analysed here was specifically scheduled to elude this critical situation, as was shown in Fig.\,\ref{Fig: Tell Corr}.

To check for these effects in previous observations, we computed the relative positions of the telluric lines during the nights analysed in \citet{delaRoche2020} and \cite{Kasper_2020}. As can be seen in Figure\,\ref{Fig: Telluric}, during the two observations presented in \citet{delaRoche2020}, the expected \ion{He}{I} absorption overlaps completely with the OH emission lines.
\citet{delaRoche2020} did not specify whether they took this telluric contamination into account and/or how they corrected the data to account for this overlap (as we did in Sect.\,\ref{Sec:Telluric}).
In such a case, it is very likely that the planetary signal is not recoverable in the transmission spectrum.
During the night reported by \citet{Kasper_2020}, the expected \ion{He}{i} absorption is clear from OH emission but overlaps with the H$_2$O absorption band. Again, imperfections in their water absorption correction have the potential to severely impact the detectability of the \ion{He}{i} absorption or its strength. These telluric contaminations, combined with the relatively low resolution of both observations, make the detection of any planetary signal extremely challenging.

Another plausible scenario able to reconcile our detection with previous null results could be the variability of the planetary \ion{He}{I} signal.
This is a case previously hinted at by \citet{Enric_2020} and \citet{Ninan2020ApJ...894...97N} in the atmosphere of GJ\,3470\,b, a planet with similar properties to GJ\,1214\,b orbiting also an M dwarf.
Long-term spectroscopic monitoring would be useful to determine whether or not GJ\,1214 presents \ion{He}{i} line variability over time.

\subsection{Stellar activity analysis}

\begin{figure}
    \centering
    \includegraphics[width=\hsize]{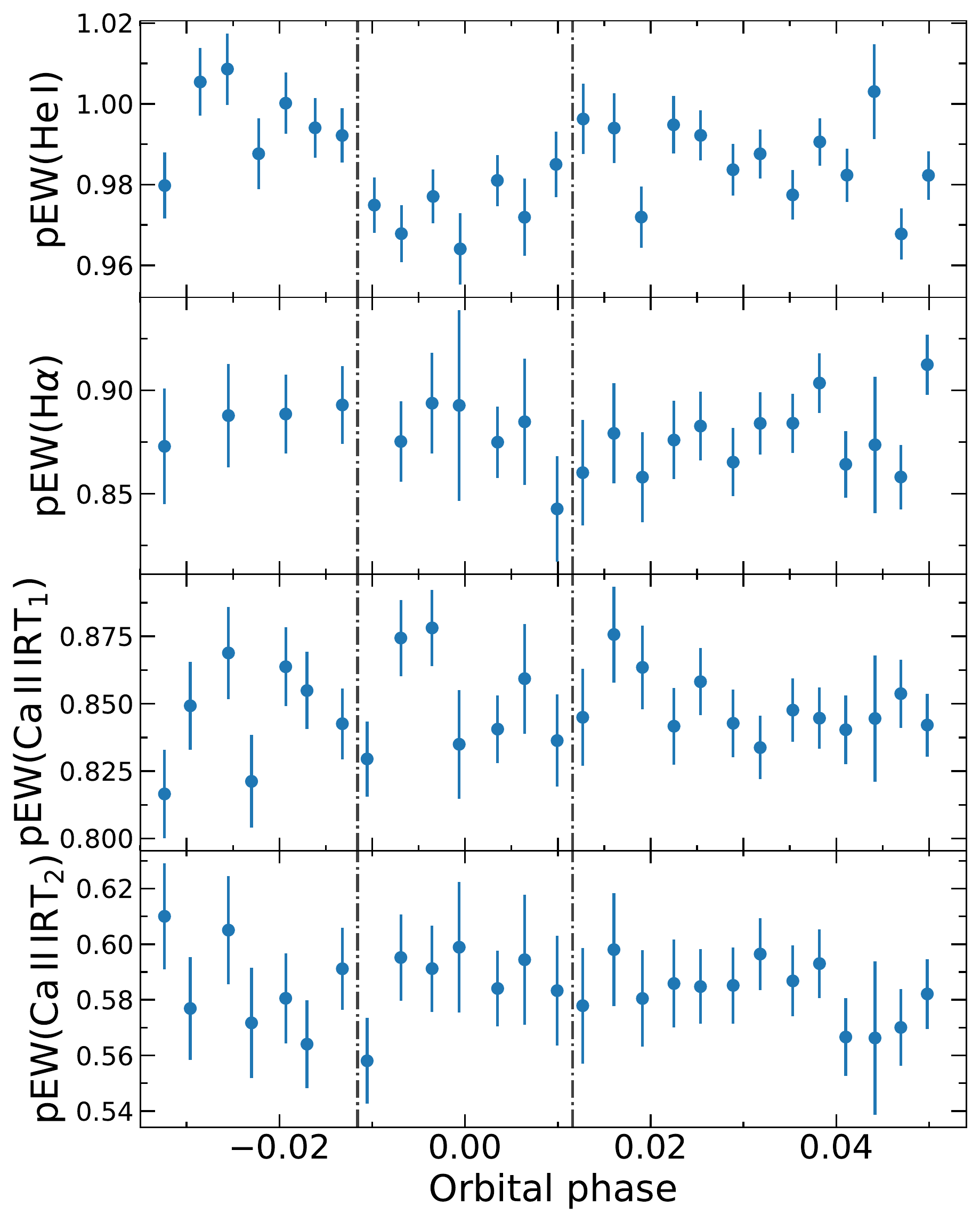}
    \caption{Time evolution of the pseudo-equivalent widths of \ion{an He}{I} triplet, H$\alpha,$ and \ion{Ca}{II}\,IRT $\lambda$8500\,$\AA$ and $\lambda$8544\,$\AA$ lines (from top to bottom).
    The vertical dashed lines mark the T1 and T4 contacts of transit of GJ\,1214\,b.
    }
    \label{Fig: pEW}
\end{figure}

It is important to take into account the presence of spots, whether or not they are occulted by the planet during the transit, because they can bias the planet parameters (\citealp{Pont_2007, Desert_2011}).
In particular, spots can change with both time and wavelength, potentially mimicking or masking signals from the planet atmosphere in transmission spectroscopy.

The GJ\,1214\,b system has been intensively observed by ground- and space-based telescopes. Long-term follow-up observations and transit observations reported the presence of starspots and active regions on the stellar surface (e.g. \citealp{Berta_2011, SysPara_Carter2011, Narita_2013, Nascimbeni_2015, Rackham_2017, Mallonn_2018}).
The insides and outsides of these regions of different brightnesses and temperatures produce 
photometric modulations detected in the long-term light curves of GJ\,1214; however, its activity pattern has not been derived precisely so far. The reported values of the stellar rotation period differ from each other: 44.3\,$\pm$\,1.2\,d (\citealp{Narita_2013}), a multiple of $\sim$\,53\,d (\citealp{Berta_2011}), $\sim$\,83\,d (\citealp{Nascimbeni_2015}), and 125\,$\pm$\,5 d \citep{Mallonn_2018}.

Stellar active regions have been shown to produce additional absorption in the \ion{He}{I} triplet lines \citep{Cauley_2018}, which are chromospheric stellar lines \citep{1986ApJS...62..899O} and thus very susceptible to stellar activity. However, the insides and outsides of active regions' dark spots or faculae (bright spots) over the visible stellar disc are unlikely to create planet-like effects during the GJ\,1214\,b transit timescale because the rotational periods reported for GJ\,1214 (P$_{\mathrm{rot}}$\,$\sim$\,40--125\,d) are considerably longer than the transit duration ($\sim$\,50\,min).
Moreover, if the planet crosses in front of dark spots during the transit, that would result in an emission signature rather than an absorption \citep{Ninan2020ApJ...894...97N}. Previous studies suggest that the photospheric activity on GJ1214 is spot dominated, with little or no evidence of faculae \citep{Rackham_2017, Mallonn_2018}.

Nevertheless, in order to study whether the \ion{He}{i} detection reported could be affected by spots or originated from stellar activity, we conducted two different analyses based on ($i$) \texttt{serval} activity indicators and ($ii$) pseudo-equivalent width (pEW) measurement.

We processed the VIS and NIR telluric-corrected spectra with \texttt{serval} \citep{SERVAL}, which is the standard CARMENES pipeline to derive the radial velocities and several activity indicators.
We looked at the evolution during the observations of the chromatic radial velocity index (CRX), the differential line width (dLW), and the H$\alpha$, \ion{Na}{I}\,D$_1$ and D$_2,$ and \ion{Ca}{II}\,IRT line indices (Fig.\,\ref{Fig: SERVAL}).
The low S/N of some orders in the VIS channel made the derivation of some line indices difficult, which can explain their large error bars and sudden variations. Overall, none of the indices exhibited evidence of variations due to stellar variability or flares. 

In a second analysis, we followed the strategy presented in \citet{Fuhrmeister_2020} to explore the variability in the stellar spectra of activity-sensitive lines.
\citet{Fuhrmeister_2020} studied the correlation between stellar activity and the \ion{He}{i} triplet variability in a sample of 319 M dwarf stars observed by CARMENES.
Their main conclusions were that 18$\%$ of the stars analysed showed \ion{He}{i} variability, which is more likely to be detected in stars with H$\alpha$ in emission, and \ion{He}{i} and H$\alpha$ line variations tend to be correlated.
Here, we focused our study on the \ion{He}{I} triplet, H$\alpha,$ and \ion{Ca}{II}\,IRT $\lambda$8500\,$\AA$ and $\lambda$8544\,$\AA$ lines. 
The low S/N around 5890\,\AA\ (S/N\,$\leq$\,1.2) prevented us from exploring the \ion{He}{I}\,D$_3,$ \ion{Na}{I}\,D$_1,$ and D$_2$ lines.
For a quantitative inspection, we computed the time evolution of the pEW for the lines of interest (see Figure\,\ref{Fig: pEW}), and we correlated them with the corrected BP$_\sigma$ TLC shown in Fig.\,\ref{Fig: TLC} ($left/blue$).
Because the \ion{He}{I} line band pass included the planetary trace, the pEW(\ion{He}{I}) exhibits an absorption feature coincident with the transit duration similar to that of the TLC (Fig.\,\ref{Fig: TLC}) and has a Pearson correlation coeficient $r=0.82$.
On the contrary, the pEW of the other lines did not exhibit line variation during the observations and do not correlate with the planetary absorption evolution, with Pearson correlation coeficients of 0.05, $-0.19,$ and $-0.21$ for H$\alpha$ and \ion{Ca}{II}\,IRT $\lambda$8500\,$\AA$ and $\lambda$8544\,$\AA$ lines, respectively.


\section{Modeling the \ion{He}{I} absorption} \label{Sec:model}



\begin{figure}
\includegraphics[width=1.0\columnwidth]{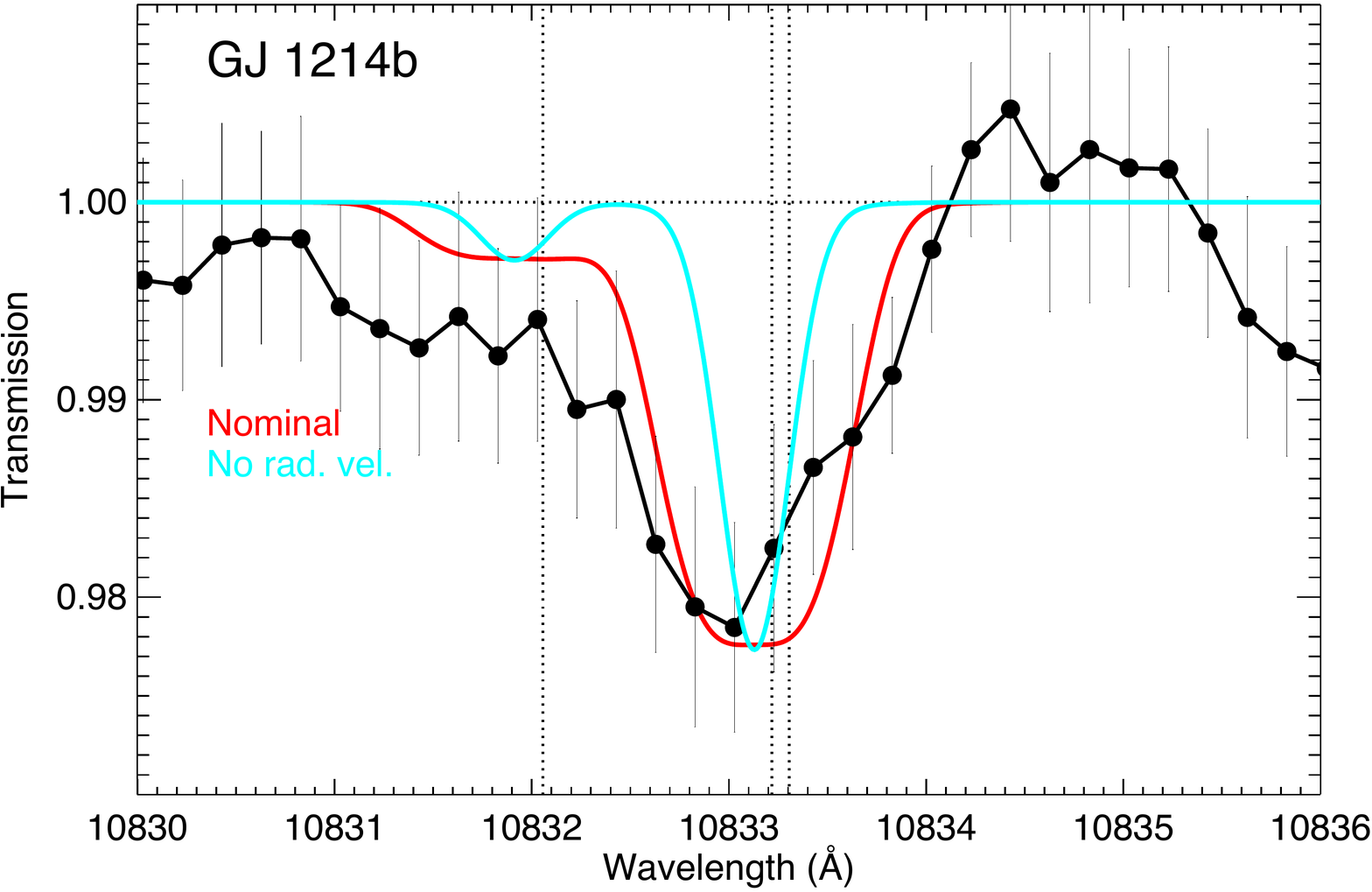}\hspace*{0.35cm}
\caption{
Phase-averaged transmission spectrum (from T1 to T4 contacts) of the He~{\sc i} triplet during transit. Measured absorption (dots), and their respective estimated errors are shown in black (as in the bottom left panel of Fig.\,\ref{Fig:MAPS_TS}), but with a two-point running mean applied\protect\footnotemark. The cyan curve shows the absorption profile when only the Doppler and turbulence broadening are included. The red curve is the profile of one of the best-fit models obtained for an effective temperature of 3600\,K, a mass-loss rate ($\dot{M}$) of 5.9$\times\,10^{10}$\,\gs,\, and an H/He mole-fraction ratio of 98/2. This absorption also includes the broadening induced by the radial velocities of the model and a blue net wind of $-4$\,km\,s$^{-1}$. The positions of the three He~{\sc i} lines are marked by vertical dotted lines.} 
    \label{fig:model}
\end{figure}
\footnotetext{The spectrum has been smoothed for illustration purposes. In the fitting analysis, the original values and errors (light grey bars in the left/bottom panel of Fig.\,\ref{Fig:MAPS_TS}) were used.}


In this section, we aim to characterise the upper atmosphere of GJ\,1214\,b from this \ion{He}{I} triplet, \het, which is a spectral absorption measurement. The analysis is based on a one-dimensional hydrodynamic and spherically symmetric model together with a non-local thermodynamic equilibrium (non-LTE) model to calculate the \het\ density distribution and the absorption profile \citep{Lampon2020}. 
In order to determine the constrained temperatures and mass-loss rates, \mlr, we followed the same method used for other exoplanets, which is described in \cite{Lampon2020} and \cite{Lampon2021a}.

Briefly, the hydrodynamic equations were
solved assuming that the escaping gas has a constant speed of sound, $v_{\textrm S}\,=\,\sqrt{k\,T_0/\bar \mu}$, where $\bar \mu$\ is the average mean molecular weight calculated in the model (see Appendix A in \citealp{Lampon2020}), and T$_0$\ is a model input parameter that is close to the maximum of the thermospheric temperature. 
We also assume that sub-stellar conditions apply to the whole planetary sphere.
Then, with the derived \het\ density distributions, we computed the phase-averaged synthetic absorption (i.e. the average from T1 to T4 contacts, as shown in Fig.\,\ref{Fig: TLC}) by using a radiative transfer code for the primary transit geometry. We recall that the helium Doppler line shapes were assumed at the atmospheric temperature T$_0$, and with an additional broadening produced by turbulent velocities. Furthermore, the component of the radial velocity of the gas along the line of sight of the observer was also included in order to account for the motion of \het\ as predicted in the hydrodynamic model. As found in the case of GJ\,3470\,b \citep{Enric_2020}, our simulations also show that the \het\ distribution of \gj\ are very extended, which, together with the fact that the size of \gj\ is also small, makes it necessary to extend the modelling of its \het\ distribution to 25\,\rp. A further extension does not significantly contribute to the absorption.

We note that, despite the large atmospheric extension of \gj, about 80\% of the \het\ phase-averaged observed  absorption takes place at altitudes below $\sim$\,8.5\,\rp, as it is limited by the subtended stellar disc. Thus, together with the fact that this absorption is rather symmetric, we do not expect significant deviations from the spherical symmetry.

For the calculation of the distribution of the \het\ densities and the radial velocity of the gas, the model requires, in addition to the planetary system parameters and the free parameters, the temperature and mass-loss rate, the stellar flux in the X-ray, and the stellar flux in the UV. The flux at $\lambda$ below 1200\,\AA\ was taken from the spectral energy distribution predicted with the coronal model discussed in Sect.\,\ref{Sec:XMM}. For the 1200--2600\,\AA\ wavelength range,\ we used the {\em Hubble}/COS (1200--1650\,\AA) and {\em Hubble}/STIS (1650--2600\,\AA) spectra. Furthermore, the model requires an H/He ratio. In our analyses of other planets, this was derived from Lyman-$\alpha$ and H$\alpha$ measurements. In this case, we are lacking of those observations and considered a rather high value of 98/2, as has been found for other planets; for example HD\,209458\,b, HD\,189733\,b, GJ\,3470\,b, and HAT-P-32\,b \citep{Lampon2020,Shaikhislamov2020,Lampon2021a,Khodachenko2021,Czesla_2021}.

We ran the model for a range of temperatures of 1000--11\,000\,K and \mlr\, of 10$^{8}$--10$^{13}$\,\gs\ in order to constrain the upper atmosphere temperature and mass loss of the planet from the measured absorption spectra. We followed the methodology described in \cite{Lampon2021a}, in particular by minimising the reduced $\chi^2$ values. Despite the large error of the measured transmission, we found a rather well-constrained range of these parameters: T\,=\,2900--4400\,K and \mlr\,=\,(1.5--18)$\times$10$^{10}$\,\gs. 

Figure~\ref{fig:model} shows the observed transmission spectrum for the corrected case (bottom left panel of Fig.\,\ref{Fig:MAPS_TS}), together with one of the best-fit spectra obtained; that for a thermospheric temperature of 3600\,K and a sub-stellar mass-loss rate of 5.9$\times10^{10}$\,\gs.
We found that the broadening of the absorption line caused by the radial expansion of the gas is critical for establishing those constraints. Figure~\ref{fig:model} shows the importance of this broadening by comparing the model absorption with those velocities and a case of no velocity. We found that at lower temperatures and \mlr\ (below 2900\,K and 1.5$\times$10$^{10}$\,\gs), the radial velocity of the gas is rather low, leading to an absorption line too narrow to fit the measured spectra. The opposite occurs at high temperatures and mass-loss rates, where the model predicts a stronger expansion producing too wide absorption spectra.

By inspection of the model results imposed by the measured spectrum, we found that the He atmosphere of this planet is very extended, producing a significant He absorption 
over the whole stellar disc during the transit. 
We also found that the gas outflow reaches the sound speed at altitudes where H atoms are in the collisional regime and overpass the Roche lobe radius at supersonic velocities. Therefore, the derived velocities from \het\ observations evidence that GJ\,1214\,b undergoes hydrodynamic escape.

Furthermore, by performing a similar analysis as that in \cite{Lampon2021b}, we found that GJ\,1214\,b is in the photon-limited hydrodynamic escape regime. Accordingly, the upper atmosphere of this planet is relatively cold, the neutral H density is dominated by advection over recombination, and the ionisation front encompasses all the escaping flow. The heating efficiency is in the range of 0.38--0.47, which shows that an important part of the absorbed XUV stellar energy is spent on driving the escape.

Comparing GJ\,1214\,b with GJ\,3470\,b, a benchmark planet for the photon-limited escape regime \citep{Lampon2021b}, we find that GJ\,1214\,b suffers a mass-loss rate of about 0.2--0.6 times that of GJ\,3470\,b, despite having almost the same gravitational potential. Moreover, GJ\,1214\,b is 2.5 times closer to its host star than GJ\,3470\,b, and the heating efficiency is up to 3.1 times higher. In contrast, the upper atmosphere of GJ\,1214\,b is colder, as the XUV stellar flux\footnote{XUV flux at $\lambda$\,<\,912\,\AA.} at planetary distance, F$_{\mathrm{XUV}}$, of GJ\,1214\,b is about six times smaller than the F$_{\mathrm{XUV}}$ received by GJ\,3470\,b. Overall, the XUV stellar irradiation is the main factor to explain mass-loss rate differences between both planets. 

Comparing our results  with the theoretical predictions of \cite{Salz_2016}, we note that their \mlr\,=\,1.9$\times$10$^{10}$\,\gs\ is in the range of mass-loss rate that we obtained. For this \mlr,\ we derived a temperature of $\approx$\,3000\,K, which is slightly higher than their maximum temperature of $\approx$\,2700\,K.
This difference in temperature can be explained by different assumptions in both models. 
While our calculated F$_{\mathrm{XUV}}$ is about 1.3 times smaller than the one used by \cite{Salz_2016}, which would lead to a lower temperature, our results yield a heating efficiency that is $\approx$\,35\%
higher, and thus give rise to a warmer temperature.
Moreover, we assumed a planetary mass that is $\sim$\,30\% higher, which also requires higher temperatures in order to reach such a mass-loss rate.
The effects of our larger gravitational potential are nevertheless ameliorated by our higher H/He ratio (98/2 versus 90/10 of \cite{Salz_2016}).

\section{Conclusions}
\label{Sec:Conclusions}

In this paper, we present the tentative detection of \ion{He}{I} $\lambda$10833\,{\AA} in the atmosphere of GJ\,1214\,b.
Using one transit observed with the CARMENES high-resolution spectrograph, we detected significant ($>$\,$4\sigma$) \ion{He}{I} excess absorption of $\sim$\,$2.1 \%$  
when corrected for telluric contamination and of 
$\sim$\,$2.2 \%$
when the contaminated regions are masked.
Furthermore, we explored the temporal behaviour of the signal by computing the \ion{He}{I} line TLC.
We detected a decrease in flux coincident with the expected GJ\,1214\,b transit duration and retrieved absorption depths consistent with the results from the transmission spectra.

The excess absorption detection reported here is in contrast with previous non-detections of the \ion{He}{I} triplet in this planet. 
We discuss how telluric contamination from both H$_2$O absorption and OH emission might have hampered the 
detections in the previous attempts, an issue that needs to be carefully considered when planning observations targeting the \ion{He}{I} triplet.
We also investigated the role of the star in the the excess absorption, but our analyses did not find any variation in the activity indices or line variations during the observations.

If confirmed by a definitive detection, this observation of \ion{He}{I} in the atmosphere of GJ\,1214\,b would be a
strong evidence that this planet has a primordial atmosphere accreted from the planetary nebula, a very important piece of information about its formation and evolution history \citep{Elkins_Tanton_2008,Kasper_2020}. In addition, it would confirm that GJ\,1214\,b is currently undergoing hydrodynamic escape, which favours the formation of a high-metallicity atmosphere and supports the results obtained by \cite{Morley_2015}. They required a high metallicity, together with the presence of clouds and hazes, in order to explain the flat transmission spectrum at optical and near-IR wavelengths of this planet.
Moreover, hydrodynamic escape is key for tracing the evolutionary path of GJ\,1214\,b. Actually, with this planet we would extend the sample of planets with measured mass-loss rates from the \ion{He}{I} triplet to smaller sizes (previously limited by GJ\,3470\,b). In that sense, if confirmed, this detection would contribute to a better understanding of planetary demography and, in particular, of the so-called Neptune desert.

Our detection of \ion{He}{I} is statistically significant at the $4.6\sigma$ level, but only one transit of GJ\,1214\,b in good telluric conditions was observed, and the repeatability of the detection could not be confirmed. Thus, further high-resolution transit observations of GJ\,1214\,b are needed to confirm (or discard) this first detection and to fully characterise the atmosphere of this benchmark sub-Neptune planet, including possible time variations. Despite the fact that GJ\,1214\,b has been intensively studied with space- and ground-based observatories, our \ion{He}{I} excess absorption is the first detection of an atmospheric species in its atmosphere.


\begin{acknowledgements}

We thank the anonymous referee for discussion and comments that helped to improve the contents of this manuscript, and Almudena Garc\'ia-L\'opez for helping to make public the NIR transmission spectrum and EUV spectra.

CARMENES is an instrument at the Centro Astron\'omico Hispano-Alem\'an (CAHA) at Calar Alto (Almer\'{\i}a, Spain), operated jointly by the Junta de Andaluc\'ia and the Instituto de Astrof\'isica de Andaluc\'ia (CSIC). 

  CARMENES was funded by the Max-Planck-Gesellschaft (MPG), 
  the Consejo Superior de Investigaciones Cient\'{\i}ficas (CSIC),
  the Ministerio de Econom\'ia y Competitividad (MINECO) and the European Regional Development Fund (ERDF) through projects FICTS-2011-02, ICTS-2017-07-CAHA-4, and CAHA16-CE-3978, 
  and the members of the CARMENES Consortium 
  (Max-Planck-Institut f\"ur Astronomie,
  Instituto de Astrof\'{\i}sica de Andaluc\'{\i}a,
  Landessternwarte K\"onigstuhl,
  Institut de Ci\`encies de l'Espai,
  Institut f\"ur Astrophysik G\"ottingen,
  Universidad Complutense de Madrid,
  Th\"uringer Landessternwarte Tautenburg,
  Instituto de Astrof\'{\i}sica de Canarias,
  Hamburger Sternwarte,
  Centro de Astrobiolog\'{\i}a and
  Centro Astron\'omico Hispano-Alem\'an), 
  with additional contributions by the MINECO, 
  the Deutsche Forschungsgemeinschaft (DFG) through the Major Research Instrumentation Programme and Research Unit FOR2544 ``Blue Planets around Red Stars'', 
  the Klaus Tschira Stiftung, 
  the states of Baden-W\"urttemberg and Niedersachsen, 
  and by the Junta de Andaluc\'{\i}a.
  
This work was based on data from the CARMENES data archive at CAB (CSIC-INTA) and made use of resources from the AstroPiso collaboration.
  
We acknowledge financial support from the Agencia Estatal de Investigaci\'on of the Ministerio de Ciencia e Innovaci\'on and the ERDF ``A way of making Europe'' through projects 
  PID2019-109522GB-C5[1:4], 
  PID2019-110689RB-I00, 
  PGC2018-098153-B-C31,
and the Centre of Excellence ``Severo Ochoa'' and ``Mar\'ia de Maeztu'' awards to the Instituto de Astrof\'isica de Canarias (CEX2019-000920-S), Instituto de Astrof\'isica de Andaluc\'ia (SEV-2017-0709), and Centro de Astrobiolog\'ia (MDM-2017-0737),
the European Research Council under the Horizon 2020 Framework Program via the ERC Advanced Grant Origins No.~832428
and via grant agreement No.~694513, 
the Excellence Cluster ORIGINS funded by the DFG under Germany's Excellence Strategy EXC-2094 No.~390783311, and the Generalitat de Catalunya via the CERCA programme.

J.O.M. agraeix el recolzament, suport i ànims que sempre ha rebut per part de padrina Conxa, padrina Merc\`e, Jeroni, Merc\`e i m\'es familiars i amics. I molt especialment a tu, Yess.The first author acknowledges the special support from Maite, Guillem, Alejandro, Benet, Joan and Montse, and the useful contributions of Andreu and Jorge. This work has made use of resources from AstroPiso collaboration.

\end{acknowledgements}

%
%

\bibliographystyle{aa}
\bibliography{references}


\begin{appendix}
\label{Sec:Appendix}

\section{Additional figures}

\begin{table}
\caption{\label{table - Priors} Parameter prior functions used in the nested sampling fit for the corrected and masked methodologies.}
\centering
\begin{tabular}{lcc}
\hline
\hline
\noalign{\smallskip}
Parameter & Corrected & Masked \\
\noalign{\smallskip}
\hline
\noalign{\smallskip}
Absorption [\%] & $\mathcal{U}(-10, 0)$ & $\mathcal{U}(-10, 0)$ \\
$\lambda_0$ [\AA] & $\mathcal{U}(10831,10835)$ & $\mathcal{U}(10831,10835)$ \\
$\sigma$ [\AA] &  $\mathcal{U}(0.01,1.0)$ & $\mathcal{N}(0.5,0.2)$ \\
\noalign{\smallskip}
\hline
\end{tabular}
\tablefoot{$^{(a)}$The prior labels $\mathcal{U}$ and $\mathcal{N}$ represent uniform and normal distributions, respectively. In the masked case, for the Gaussian standard deviation ($\sigma$), we used a normal prior to avoid unrealistic Gaussian solutions due to the lack of points. }
\end{table}

\begin{figure}
    \centering
    \includegraphics[width=\hsize]{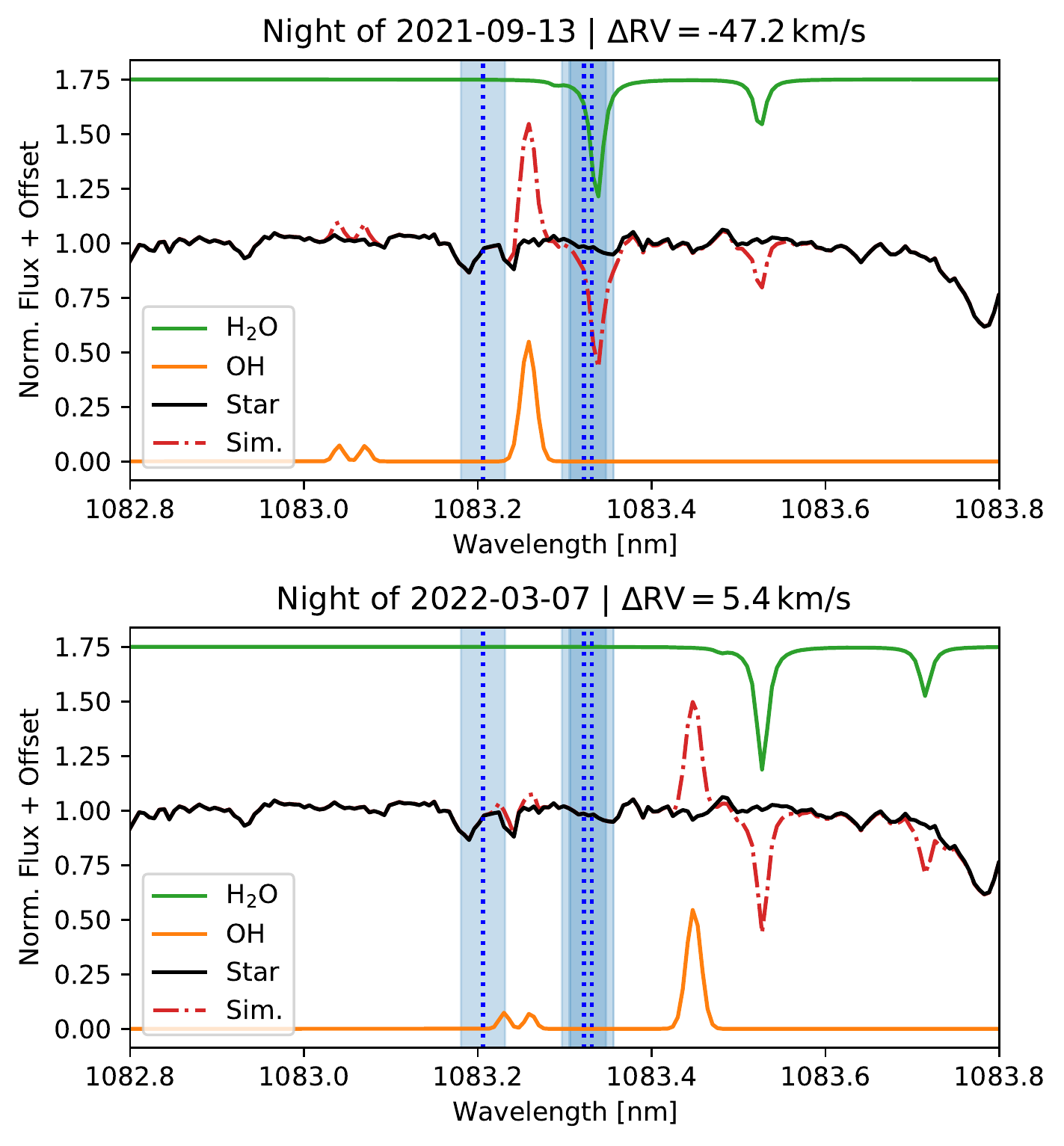}
    \caption{Same as Fig.\,\ref{Fig: Telluric}, but for two nights with the bluest ($top$) and reddest ($bottom$) shifts of the telluric lines with respect to the \ion{He}{I} lines.
    }
    \label{Fig: MIN_MAX_OH_position}
\end{figure}

\begin{figure}
     \centering
     \begin{subfigure}{0.49\textwidth}
         \centering
         \includegraphics[width=\textwidth]{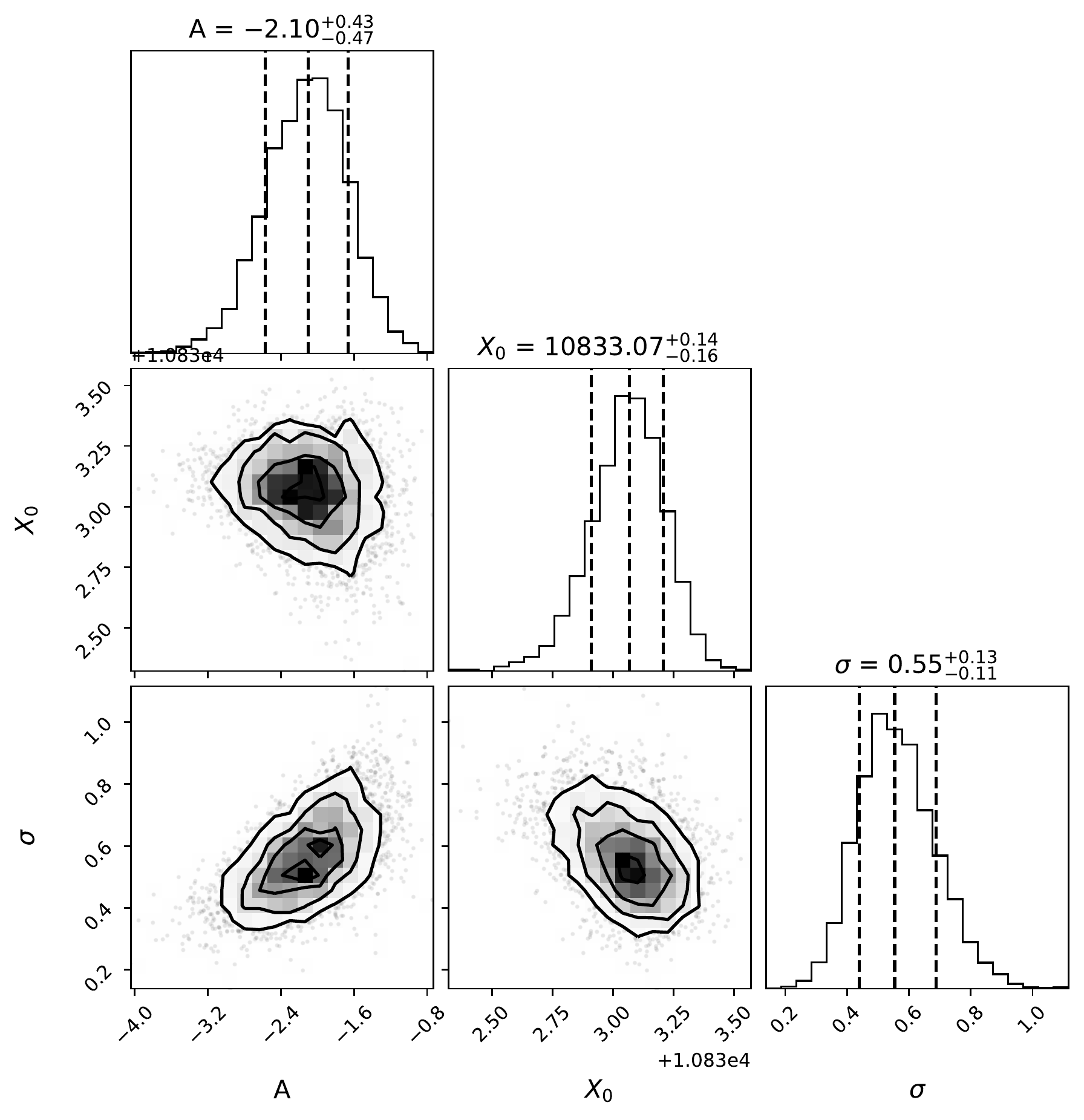}
     \end{subfigure}
     \hfill
     \begin{subfigure}{0.49\textwidth}
         \centering
         \includegraphics[width=\textwidth]{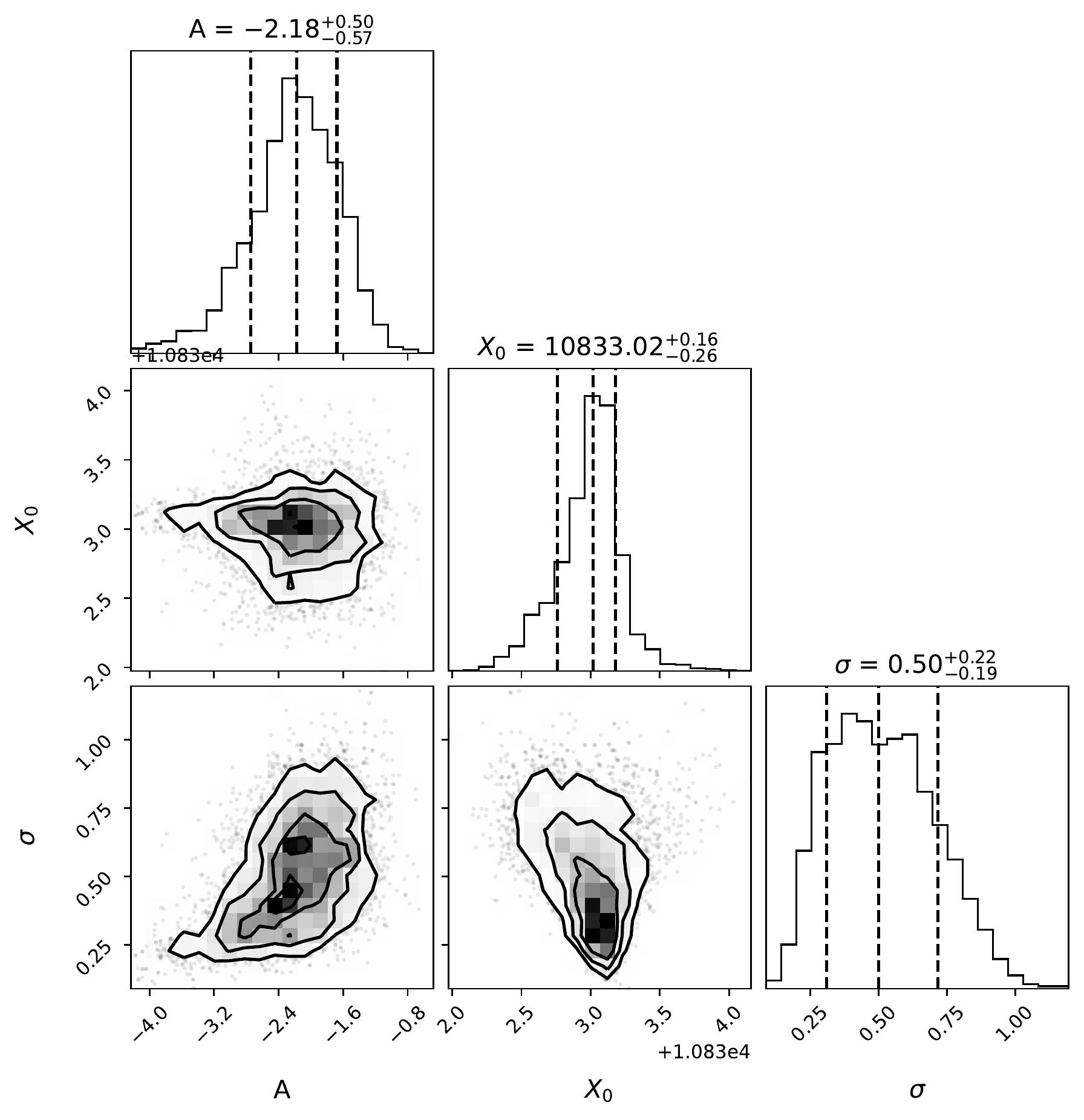}
     \end{subfigure}
        \caption{Corner plot for the nested sampling posterior distribution of the Gaussian profile parameters for the corrected (\textit{top}) and masked (\textit{bottom}) procedures, prepared with the \texttt{corner.py} package (\citealp{Corner_plot}).
        }
        \label{Fig: Corner plot}
\end{figure}

\begin{figure}
    \centering
    \includegraphics[width=\hsize]{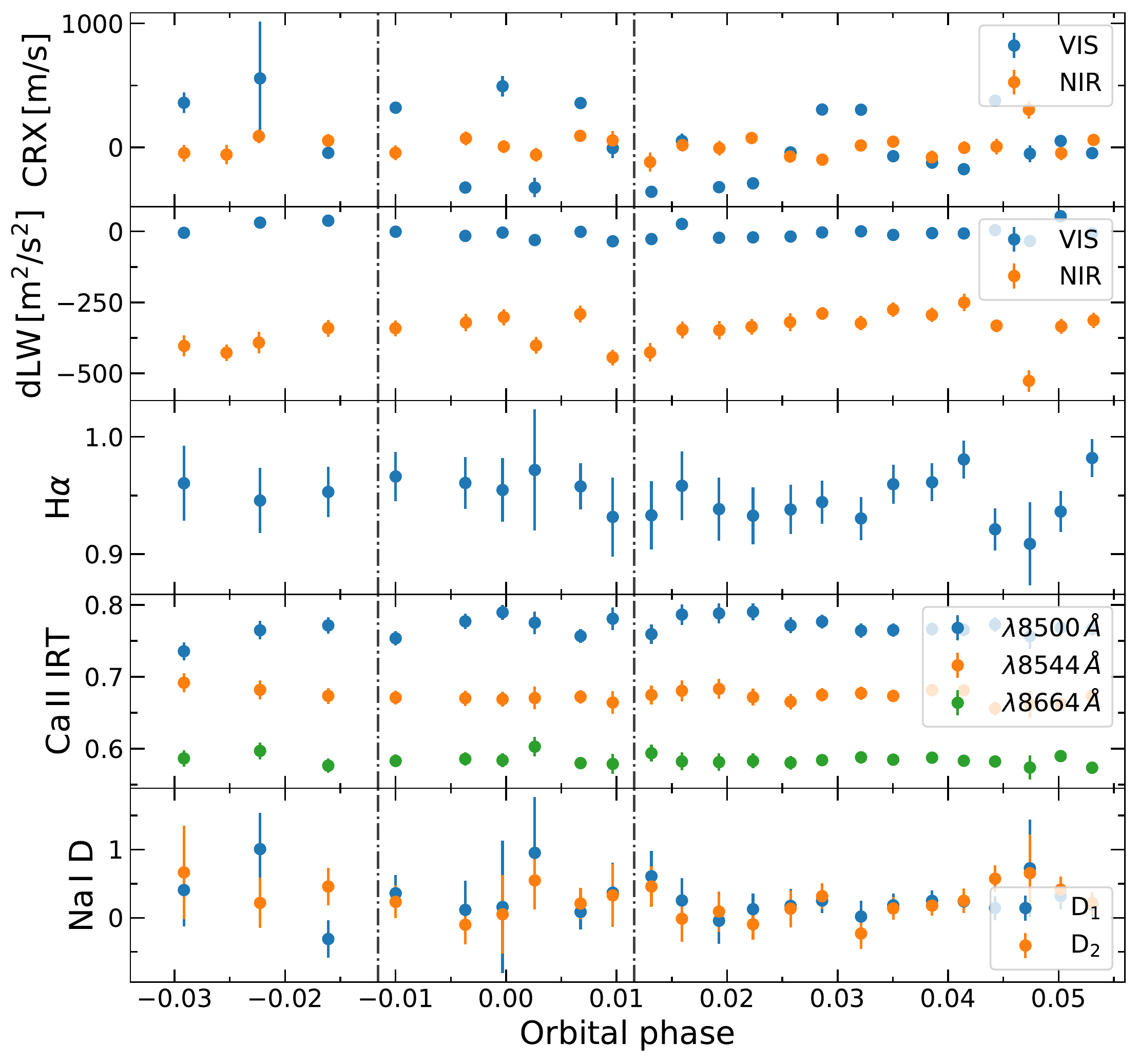}
    \caption{Activity indices derived using \texttt{serval}. From top to bottom: chromatic radial velocity index (CRX), diferential line width (dLW), H$\alpha$ line index; \ion{Ca}{II}\,IRT line indices, and \ion{Na}{I}\,D$_1$ and D$_2$ line indices. The vertical dashed lines mark the $T_1$ and $T_4$ contacts of transit of GJ\,1214\,b.
    }
    \label{Fig: SERVAL}
\end{figure}

\section{Molecfit parameters}

\begin{table}
\caption{\label{table - molecfit parameters} Initial \url{molecfit} fit parameters.}
\centering
\begin{tabular}{lcl}
\hline
\hline
\noalign{\smallskip}
Parameter & Value & Description \\
\noalign{\smallskip}
\hline
\noalign{\smallskip}
{\tt ftol} & $10^{-2}$ & Relative $\chi ^2$ convergence criterion \\
{\tt xtol} & $10^{-2}$ & Relative parameter convergence criterion \\
{\tt list\_molec} & H$_2$O, O$_2$, CH$_4$, CO$_2$ & Fitted molecules \\
{\tt cont\_n} & 3 & Degree of polynom for continuum fit \\
{\tt cont\_const} & 1 & Constant of polynom for continuum fit \\
{\tt wlc\_n} & 1 & Degree of Chebyshev polynom for wavelength calibration fit \\
{\tt wlc\_const} & 0 & Initial constant of Chebyshev polynom for wavelength calibration fit \\
{\tt res\_gauss} & 5.25196 & FWHM of Gaussian in pixels (fixed) \\
{\tt res\_lorentz} & 0.75664 & FWHM of Lorentzian in pixels (fixed) \\
{\tt kernfac} & 18 & Kernel size \\
{\tt varkern} & 1 & Linear increase of kernel with wavelength \\ 
{\tt ref\_atm} & ngt & Reference MIPAS atmospheric profile \\
{\tt layers} & 1 & Fixed grid \\
{\tt emix} & 5 & Upper mixing high \\
{\tt pwv} & -1 & No scaling of input water vapour profile \\
\noalign{\smallskip}
\hline
\end{tabular}
\end{table}

\begin{table}
\caption{\label{table - molecfit ranges} Fitted wavelength regions in \url{molecfit}.}
\centering
\begin{tabular}{c}
\hline
\hline
\noalign{\smallskip}
$\Delta \lambda$ [$\mu$m] \\
\noalign{\smallskip}
\hline
\noalign{\smallskip}
0.9635--0.9855 \\
1.0710--1.1111 \\
1.1640--1.2320 \\
1.2500--1.2820 \\
1.2880--1.3110 \\
1.5680--1.5800 \\
1.5980--1.6160 \\
1.6360--1.6640 \\
\noalign{\smallskip}
\hline
\end{tabular}
\end{table}

\end{appendix}

\end{document}